\documentclass[twocolumn]{emulateapj}

\usepackage{times}

\usepackage{amsmath,amssymb,amstext}
\usepackage[colorlinks=blue,linkcolor=blue,urlcolor=blue,citecolor=blue]{hyperref} 
\usepackage[all]{hypcap}
\usepackage{tabularx}
\usepackage{float}
\usepackage{natbib}
\usepackage[usenames, dvipsnames]{color}

\newcommand{\kms}{km\,s$^{-1}$}

\begin{document}
\title{EVIDENCE FOR THE MAGNETIC BREAKOUT MODEL IN AN EQUATORIAL CORONAL-HOLE JET}
\author{Pankaj Kumar\altaffilmark{1}\footnote{NPP Fellow}, Judith T. Karpen\altaffilmark{1}, Spiro K. Antiochos\altaffilmark{1}, Peter F. Wyper\altaffilmark{2}, C. Richard DeVore\altaffilmark{1}, Craig E. DeForest\altaffilmark{3}}
\affil{$^1$Heliophysics Science Division, NASA Goddard Space Flight Center, Greenbelt, MD, 20771, USA}
\affil{$^2$Department of Mathematical Sciences, Durham University, Durham DH1 3LE, UK}
\affil{$^3$Southwest Research Institute, 1050 Walnut Street, Boulder, CO, USA}
\email{pankaj.kumar@nasa.gov}

\begin{abstract}
Small, impulsive jets commonly occur throughout the solar corona, but are especially visible in coronal holes. Evidence is mounting that jets are part of a continuum of eruptions that extends to much larger coronal mass ejections and eruptive flares. Because coronal-hole jets originate in relatively simple magnetic structures, they offer an ideal testbed for theories of energy buildup and release in the full range of solar eruptions. We analyzed an equatorial coronal-hole jet observed by SDO/AIA on 09 January 2014, in which the magnetic-field structure was consistent with the embedded-bipole topology that we identified and modeled previously as an origin of coronal jets. In addition, this event contained a mini-filament, which led to important insights into the energy storage and release mechanisms. SDO/HMI magnetograms revealed footpoint motions in the primary minority-polarity region at the eruption site, but show negligible flux emergence or cancellation for at least 16 hours before the eruption. Therefore, the free energy powering this jet probably came from magnetic shear concentrated at the polarity inversion line within the embedded bipole. We find that the observed activity sequence and its interpretation closely match the predictions of the breakout jet model, strongly supporting the hypothesis that the breakout model can explain solar eruptions on a wide range of scales. 
\end{abstract}

\section{INTRODUCTION}
Solar jets are transient plasma ejections that occur repeatedly in coronal holes, quiet corona, and active regions, and may supply a significant amount of mass and energy to the corona and solar wind \citep{raouafi2016}. Most previous studies of coronal-hole (CH) jets only addressed those events occurring in polar holes and their evolving properties derived from extreme ultraviolet/soft X-ray (EUV/SXR) images. Because magnetograms near the limb are of poor quality, the underlying magnetic-field properties could not be determined. The speeds, lifetimes, and other physical properties of several polar CH jets were derived from Hinode/XRT observations \citep{savcheva2007,cirtain2007}. Polar CH jets frequently exhibit helical structure and untwisting motions  \citep{patsourakos2008,moore2015}. \citet{nistico2009} studied 79 jets in polar CHs using STEREO observations, and classified them structurally in terms of Eiffel-tower, lambda, and micro-CME type jets. \citet{raouafi2010} found an association between X-ray jets and S-shaped micro-sigmoids, and suggested that micro-sigmoids may be progenitors of coronal jets.

In contrast, studies of equatorial coronal-hole jets are rare. For example, \citet{nistico2010} reported the observational features of 15 equatorial CH jets using STEREO observations, and found no significant physical difference between equatorial and polar CH jets. Their average speed and duration were found to be $\sim$200 \kms and 30 min, respectively. A recent study of 20 polar CH jets found that most, if not all, of these events were triggered by mini-filament eruptions \citep{sterling2015}. However, this study could not determine the trigger or the magnetic configuration due to the lack of photospheric magnetic-field data near the poles. 

Here we present the analysis and interpretation of an on-disk jet in an equatorial coronal hole, for which we could observe the magnetic field evolution and determine the most likely trigger/driver. Our previous numerical studies of reconnection-driven coronal jets identified a fundamental magnetic-field topology -- the embedded bipole -- as well as a mechanism of energy buildup and explosive release that yields Alfv\'enic, helical outflows consistent with observations \citep{pariat2009,pariat2010,pariat2015,pariat2016,wyper2016a,wyper2016b,karpen2017}.  
 \citet{wyper2017} demonstrated that our breakout model for large-scale solar eruptions also explains small-scale jets. In contrast to our previous studies, this variant of the embedded-bipole paradigm also produces a mini-filament eruption, in agreement with the \citet{sterling2015} observations.  In this paper, we report observations and analysis of a well-observed equatorial coronal-hole jet that closely agree with the predictions of the breakout jet model \citep{antiochos1998,antiochos1999}. We present the observations in \S \ref{obs}; \S \ref{model} briefly reviews the key features of our embedded-bipole jet model with and without filament eruptions; \S \ref{results} describes our interpretation of the observed event.  In \S5, we summarize our conclusions regarding the pre-event configuration, the key points of agreement with the breakout jet simulations, and evidence for the breakout mechanism in this case.
\begin{figure*}
\centering{
\includegraphics[width=11cm]{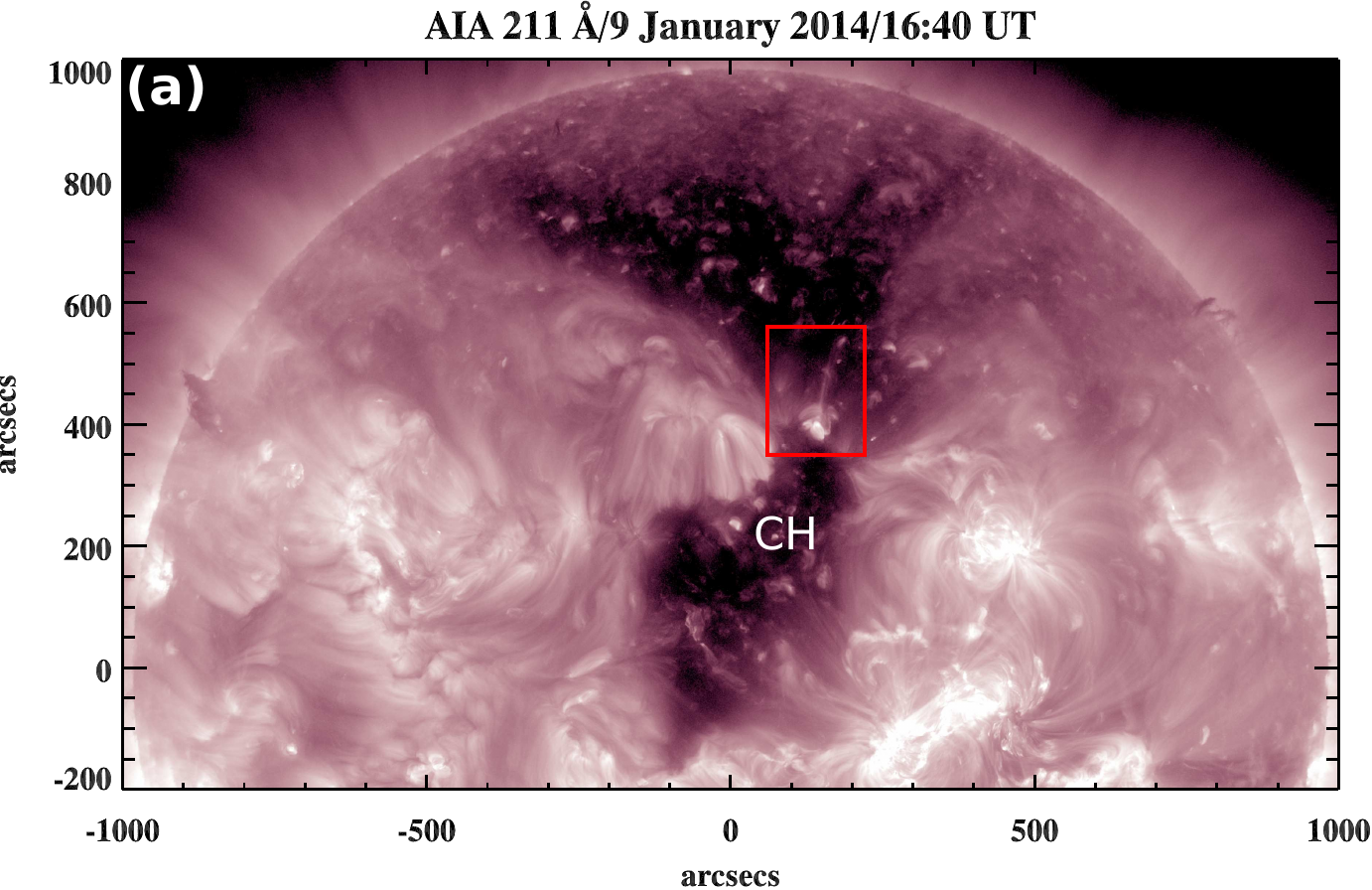}
\includegraphics[width=6cm]{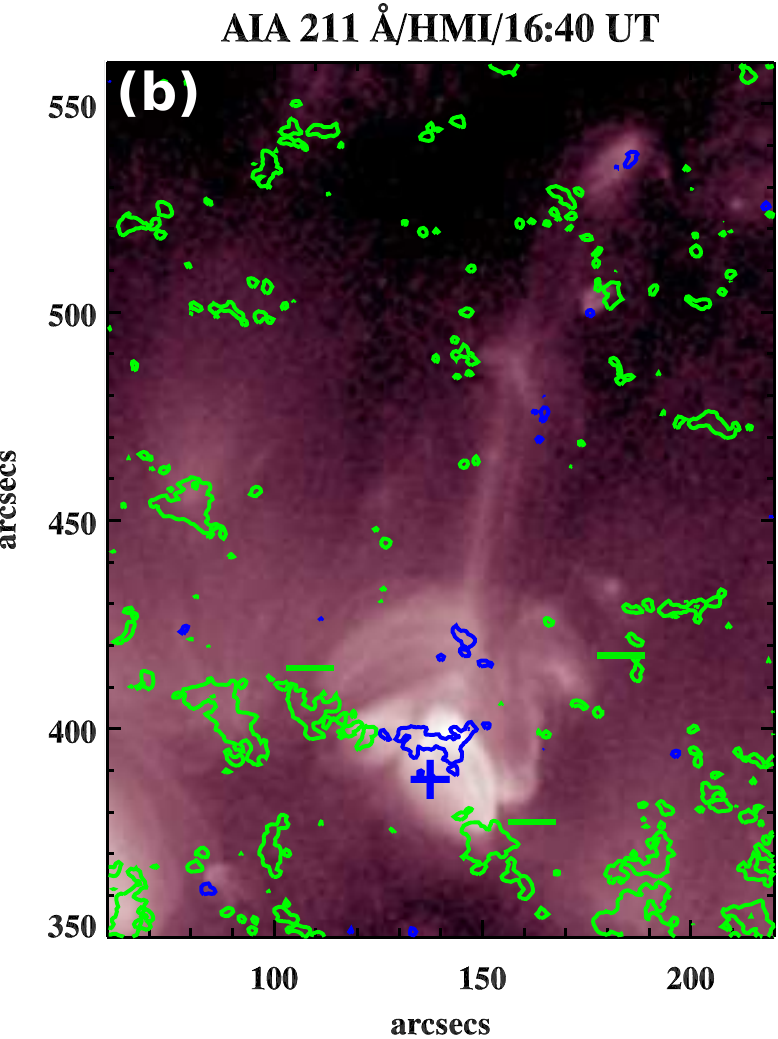}
}
\caption{(a) AIA 211 \AA~ image showing the CH containing the jet source region marked by a red box. (b) Enlarged view of the jet source region (red box in (a)). HMI magnetogram contours ($\pm$50 Gauss) of positive (blue) and negative (green) polarities are superposed on the EUV image. 
} 
\label{aia211}
\end{figure*}
\begin{figure*}
\centering{
\includegraphics[width=16cm]{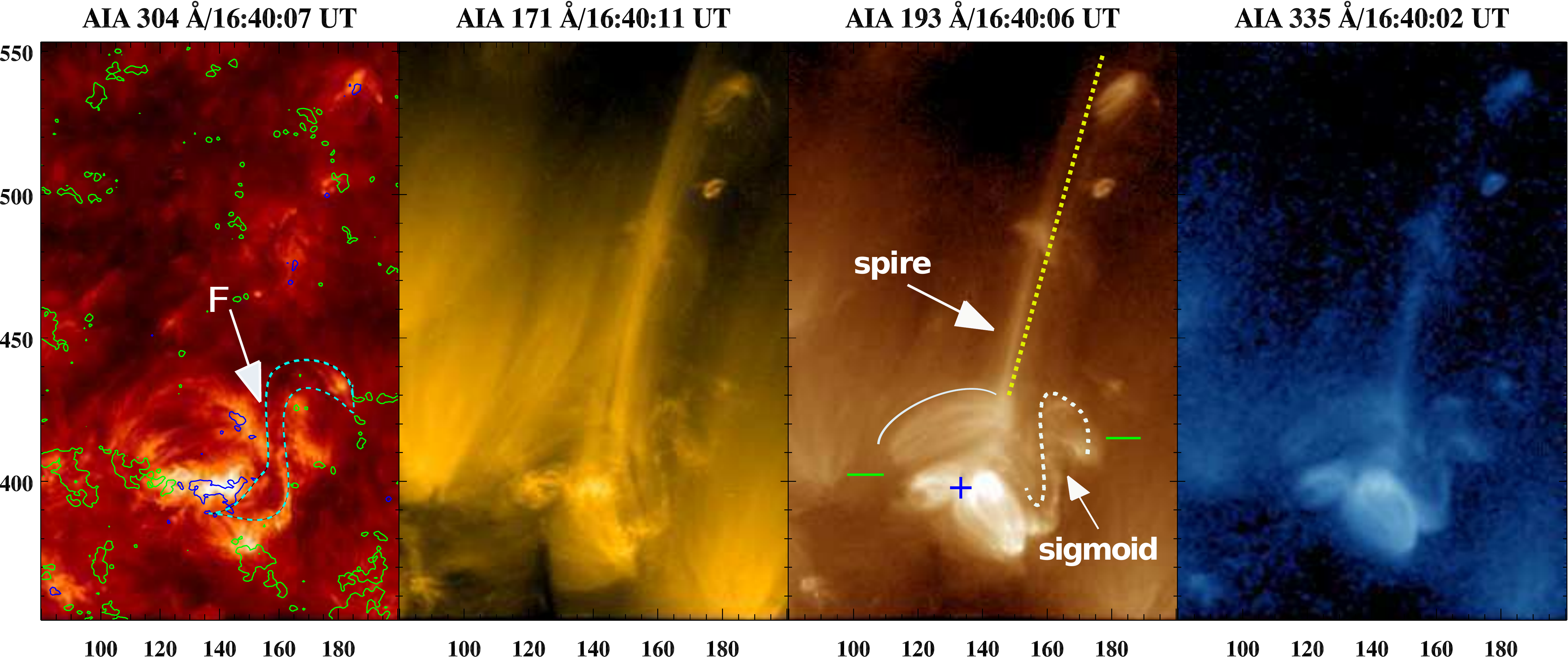}
\includegraphics[width=16cm]{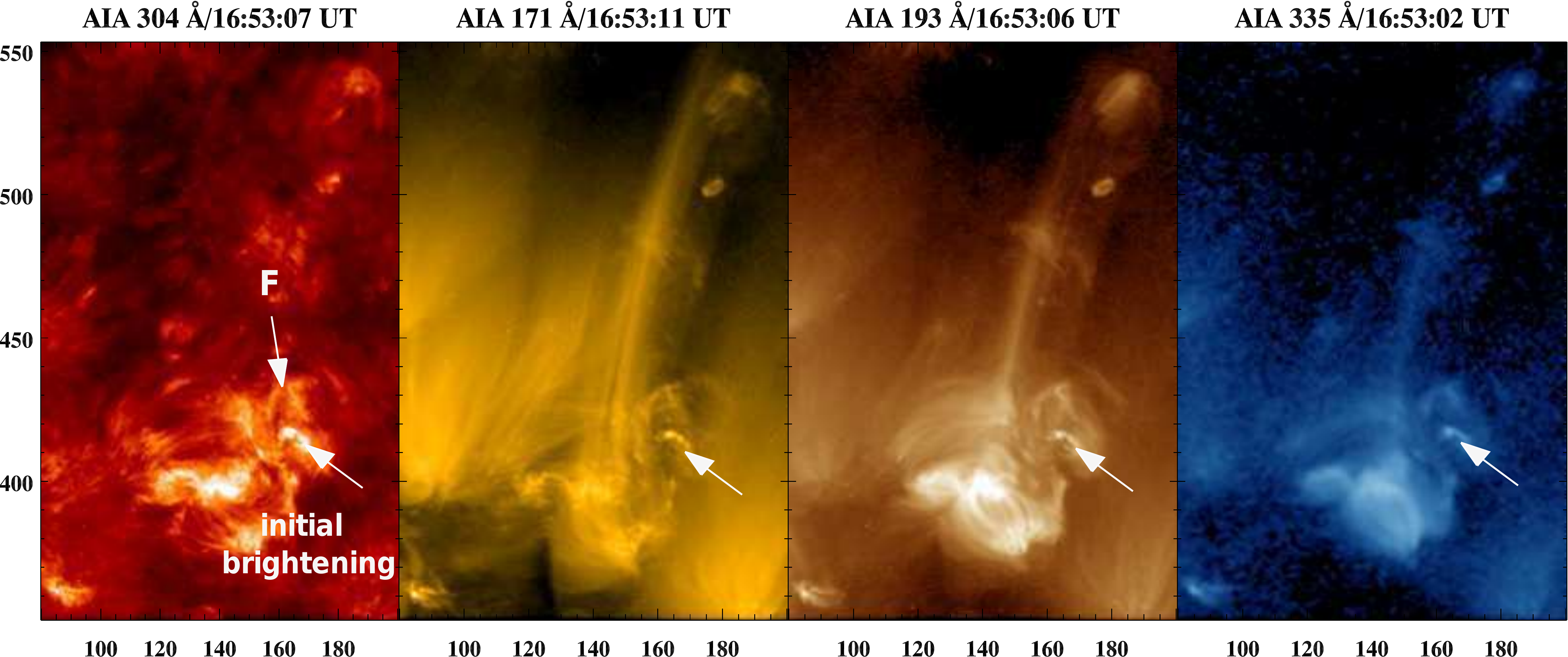}
\includegraphics[width=16cm]{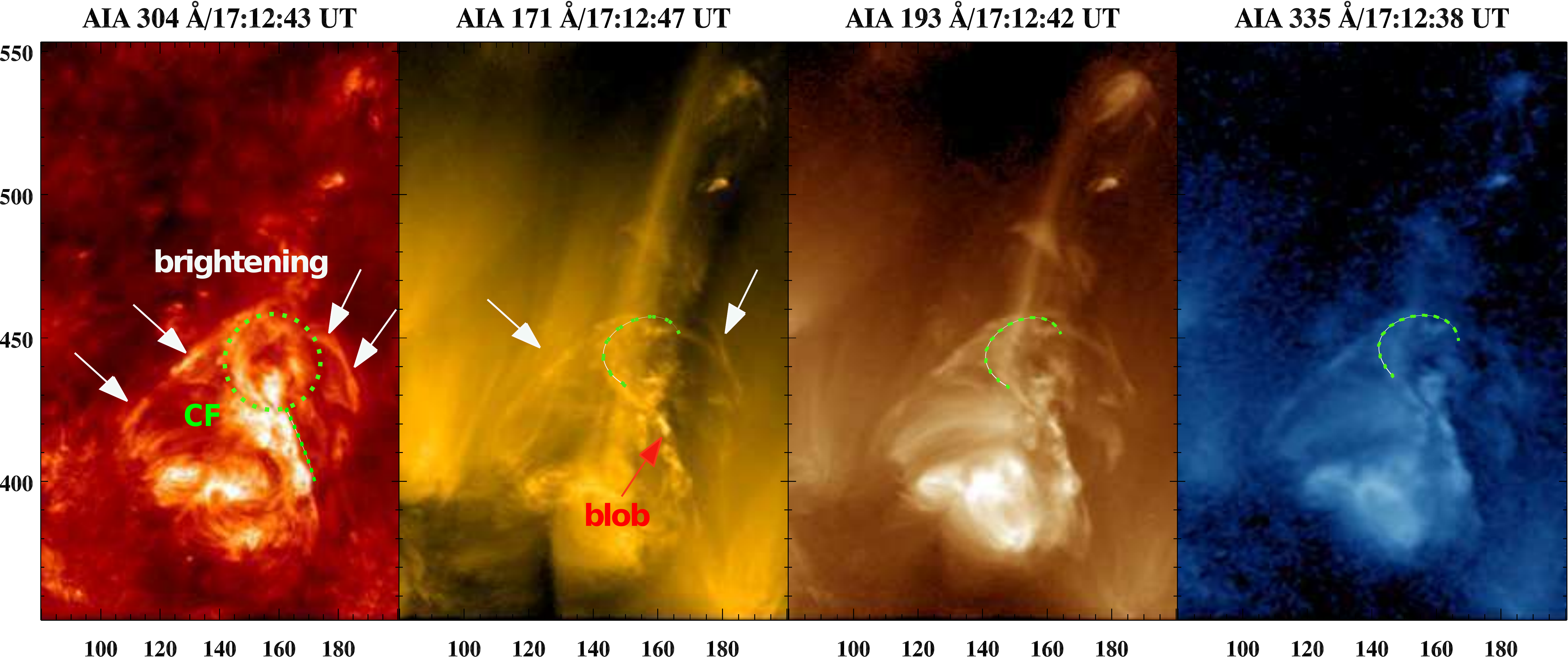}
}
\caption{Selected AIA 304, 171, 193, and 335 \AA~ images showing the CH jet source region at selected times before jet onset. The earliest AIA 304 \AA~ image is overlaid by HMI magnetogram contours ($\pm$50 Gauss) of positive (blue) and negative (green) polarities. The mini-filament is marked by a white arrow labeled F and a dashed outline in the same panel. The spire (yellow dashed line), one side of the fan (solid white arc), and sigmoid (white dashed line) are marked in the earliest 193\AA~ panel.  The green dotted line in the bottom panels bounds the circular feature (CF) surrounding the filament. Arrows point to various bright features as labeled and discussed in the text. X- and Y axes are labeled in arcsecs. The full dynamic evolution of this region is shown in the accompanying movie. 
} 
\label{aia1}
\end{figure*}
\begin{figure*}
\centering{
\includegraphics[width=4.3cm]{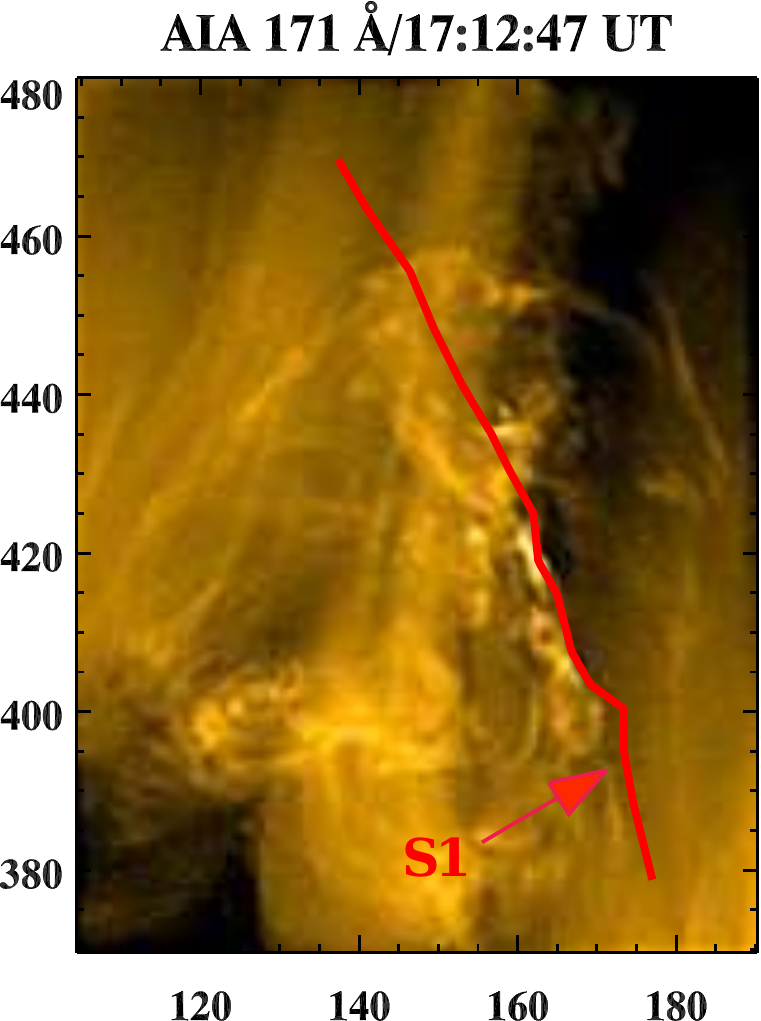}
\includegraphics[width=4.3cm]{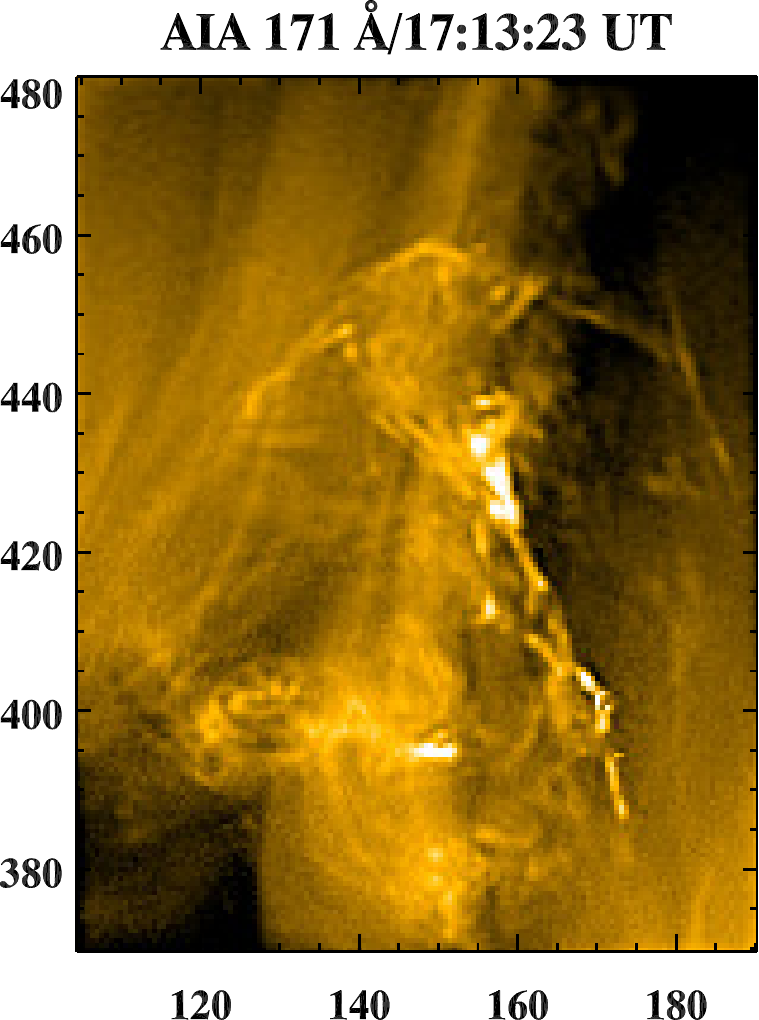}
\includegraphics[width=4.3cm]{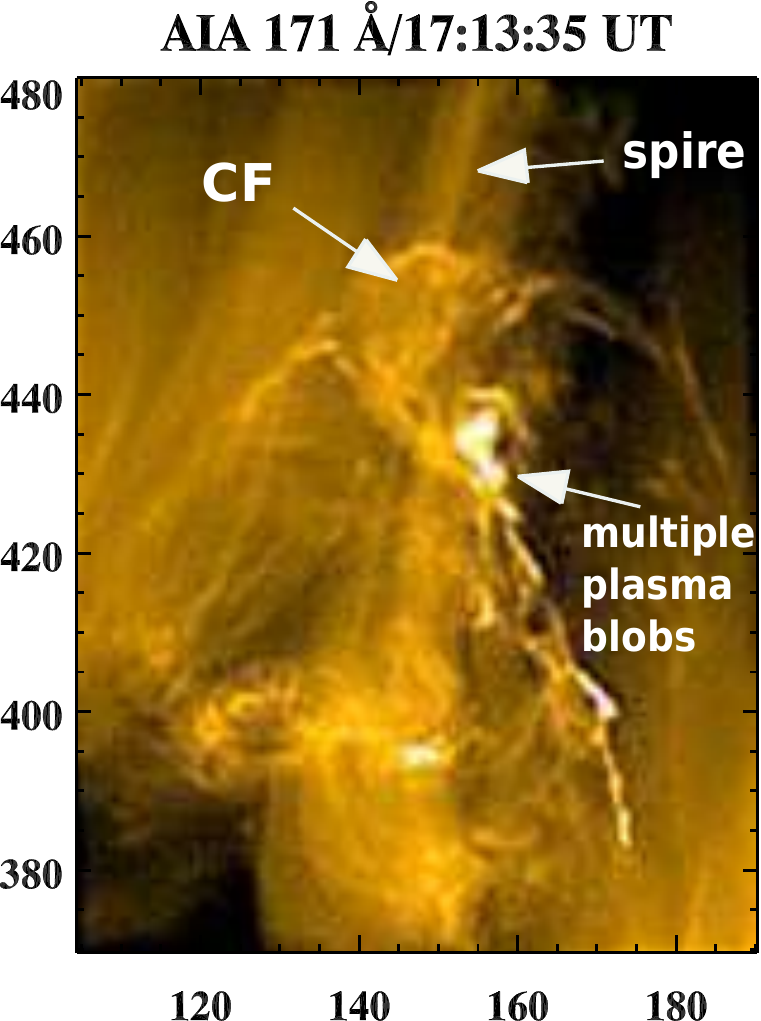}
\includegraphics[width=4.3cm]{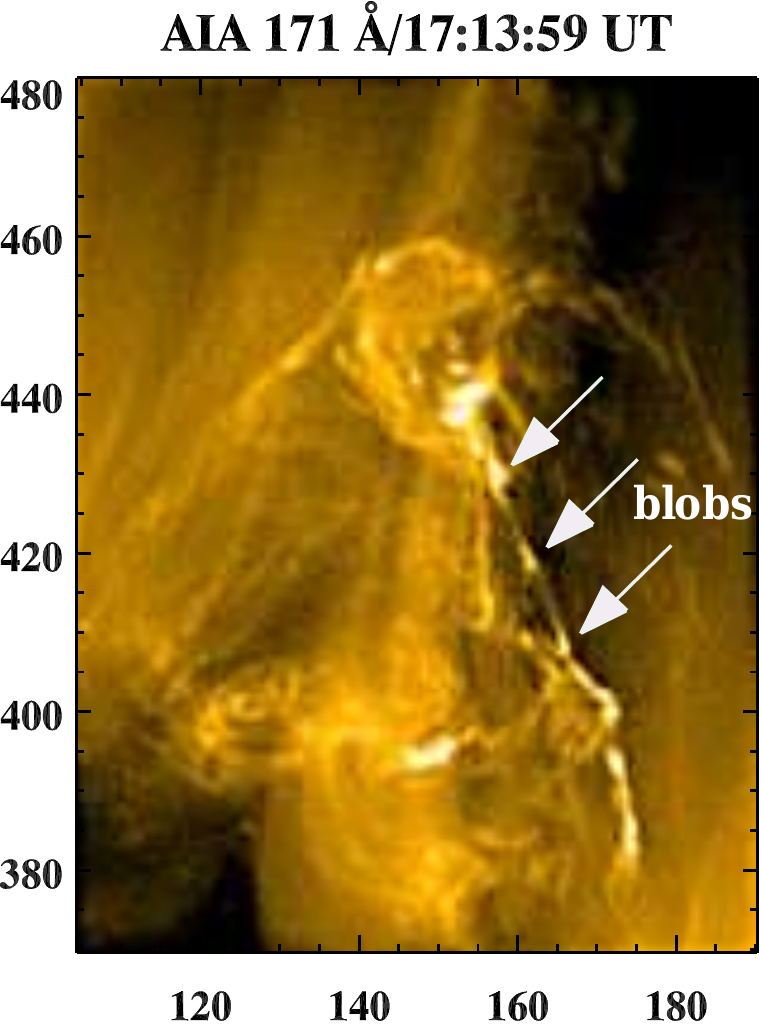}

\includegraphics[width=4.3cm]{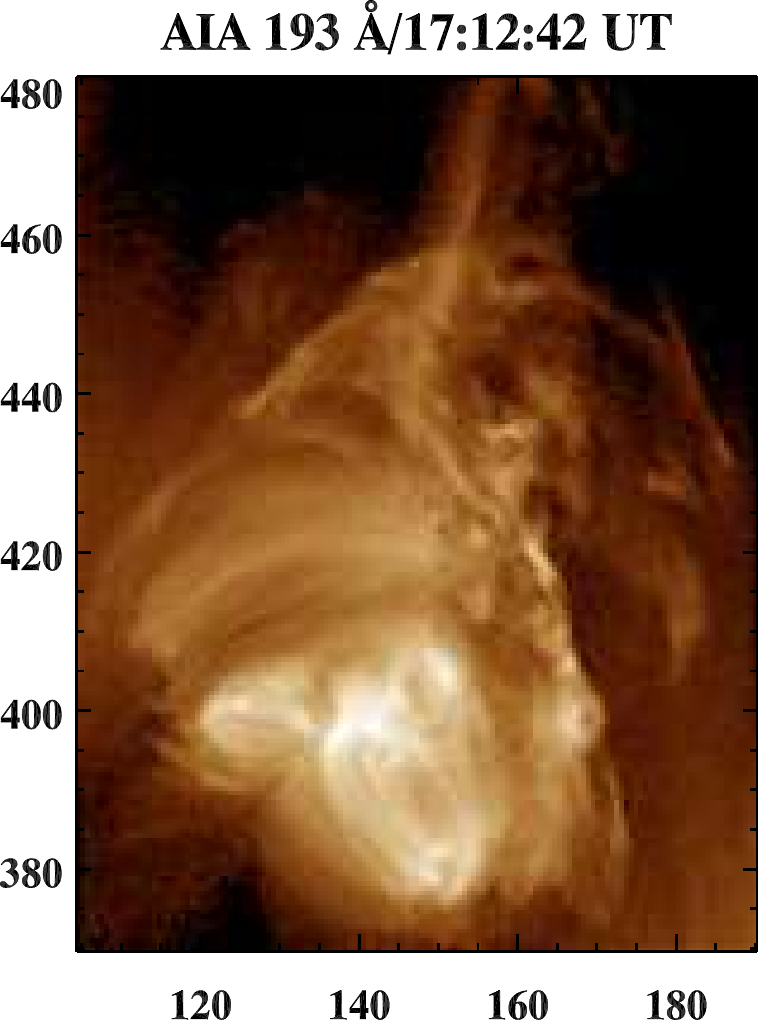}
\includegraphics[width=4.3cm]{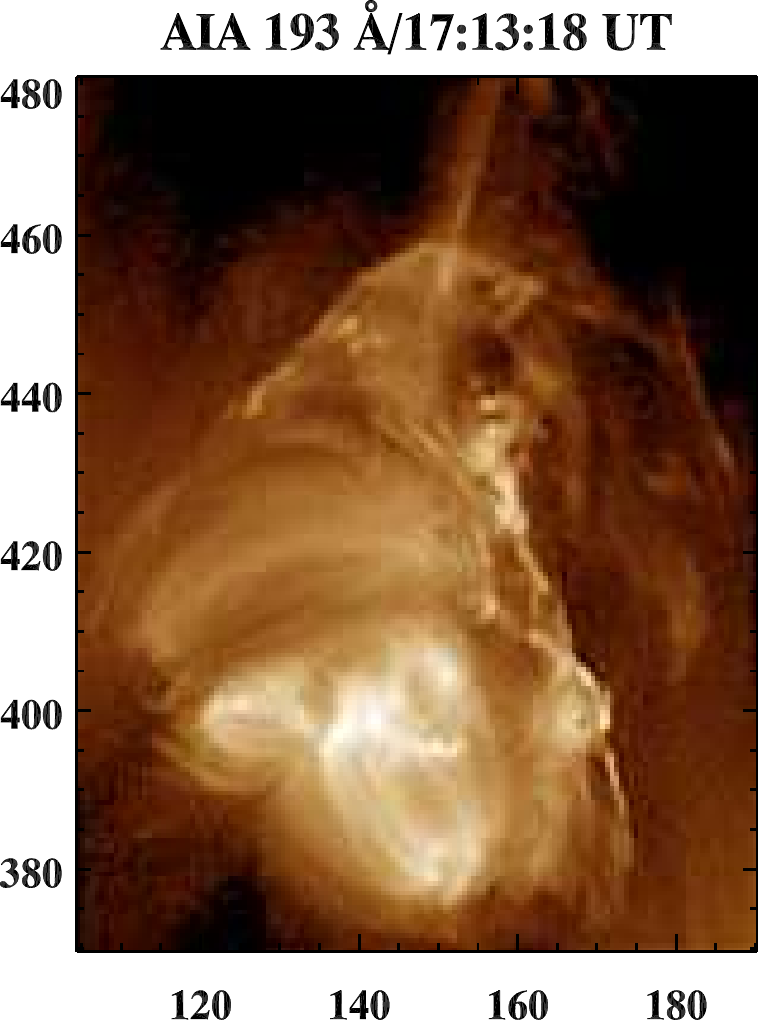}
\includegraphics[width=4.3cm]{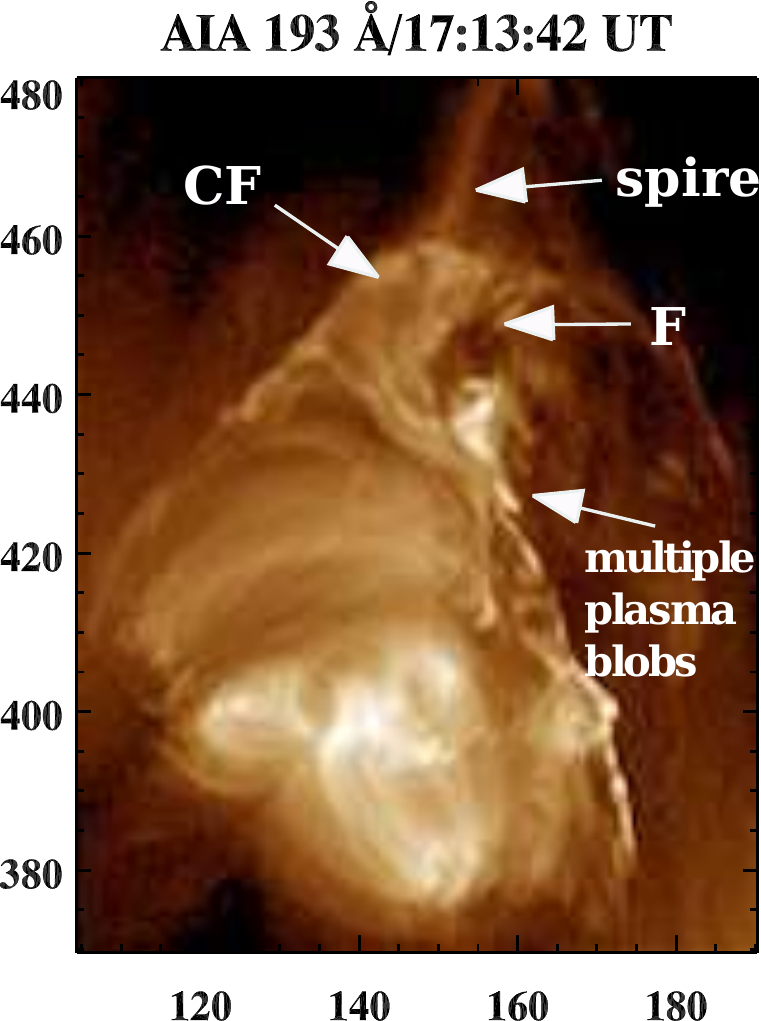}
\includegraphics[width=4.3cm]{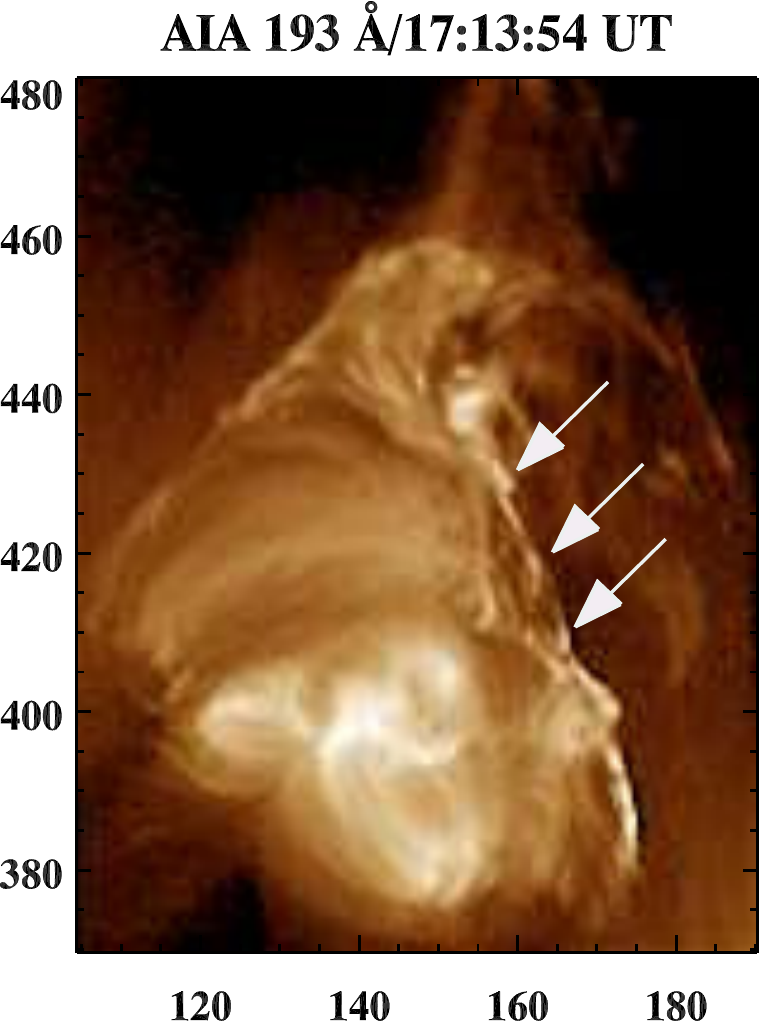}

\includegraphics[width=4.3cm]{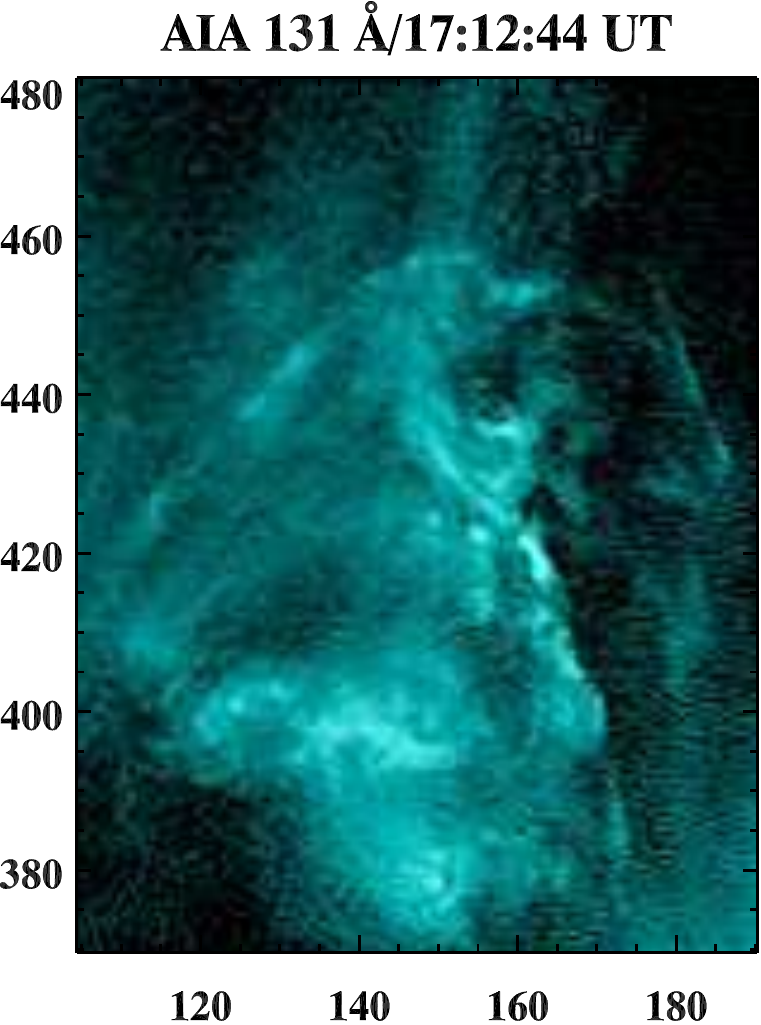}
\includegraphics[width=4.3cm]{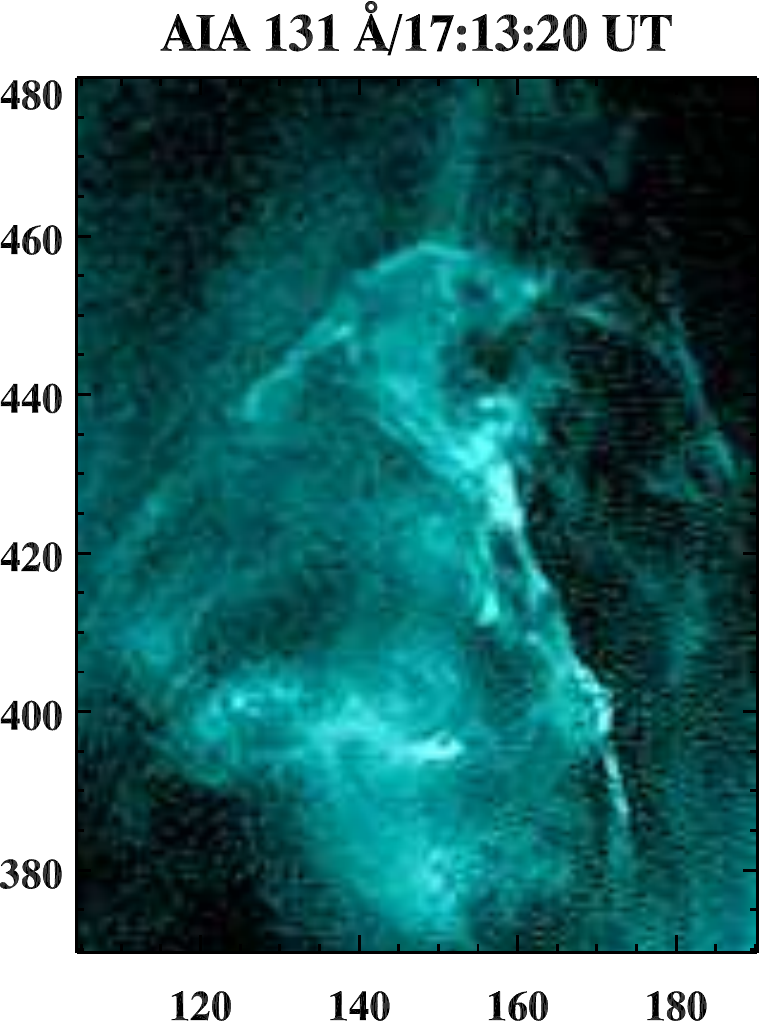}
\includegraphics[width=4.3cm]{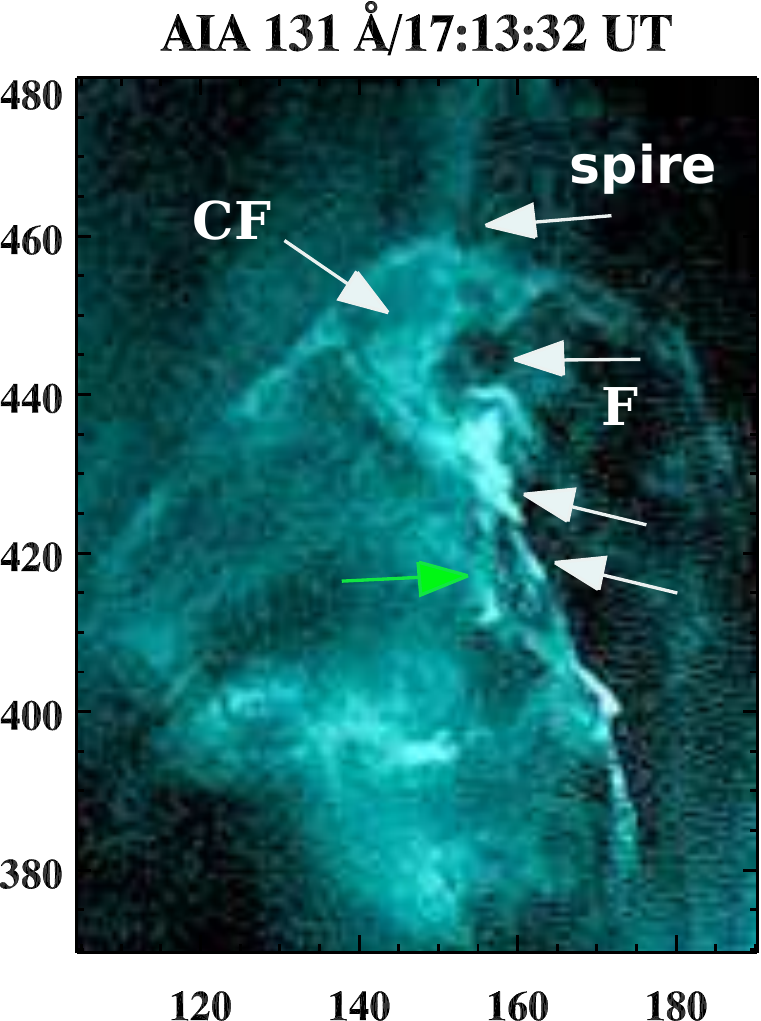}
\includegraphics[width=4.3cm]{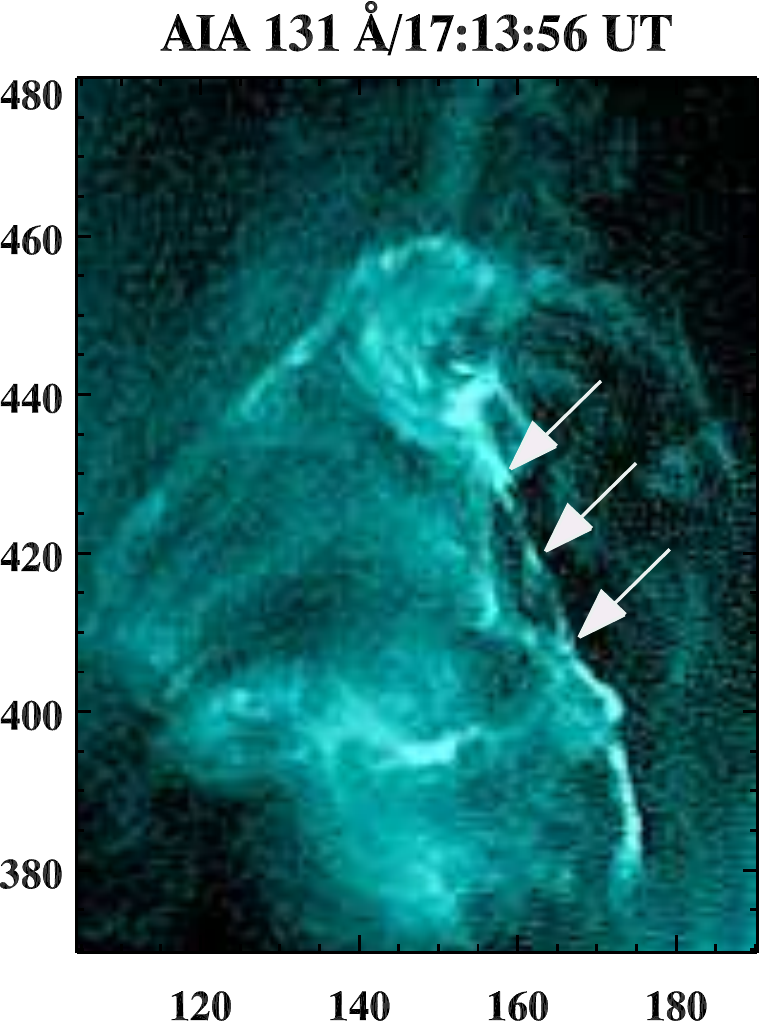}
}
\caption{Selected AIA 171, 193, and 131 \AA~ images showing multiple hot plasma blobs behind the rising circular feature (CF) containing the dark mini-filament (F). The red line labelled S1 in the top left panel is the slice used to create the time-distance intensity plot in Figure \ref{st}(a-c). The green arrow in the 131 \AA~ panel at 17:13:32 UT points to the inverted-V shaped structure discussed in \S\S \ref{obs} and \ref{results}, which is clearly visible at all times and wavelengths shown here. X and Y axes are labeled in arcsecs. The accompanying movie shows the full dynamic evolution from 17:12:42 to 17:13:59 UT. 
} 
\label{blobs}
\end{figure*}
\begin{figure}
\centering{
\includegraphics[width=8.9cm]{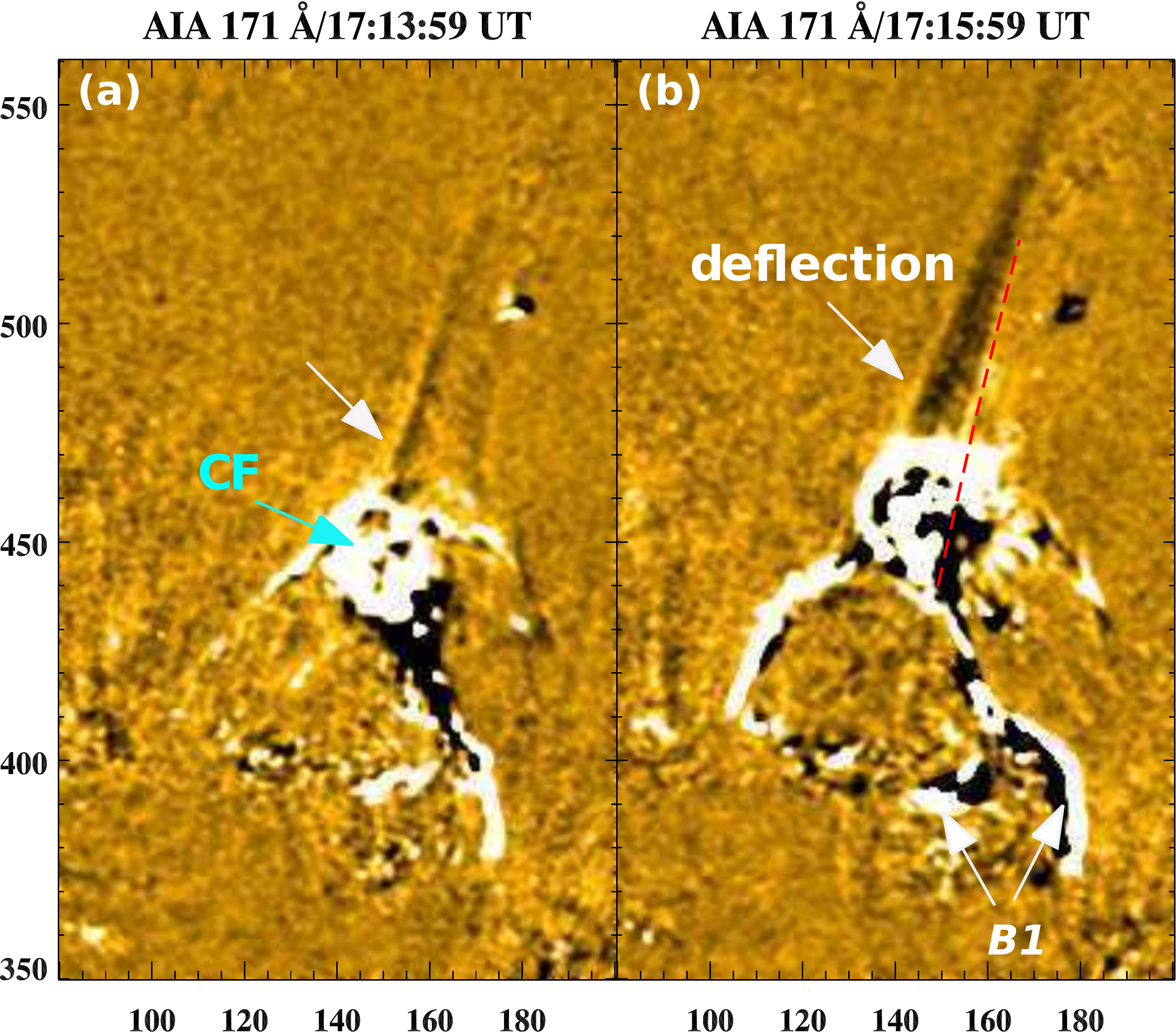}
}
\caption{(a-b) AIA 171 \AA~ running difference ($\Delta$t=1 min) images at 2 selected times before jet onset, showing the progressive deflection of the spire, the rising fan surface, and the circular feature (CF).  The red dashed line shows the orientation and extent of the spire at 16:40 UT. X and Y axes are labeled in arcsecs. The full temporal evolution from 16:40 UT to 17:30 UT is shown in the accompanying movie. } 
\label{def}
\end{figure}
\begin{figure*}
\centering{
\includegraphics[width=5.5cm]{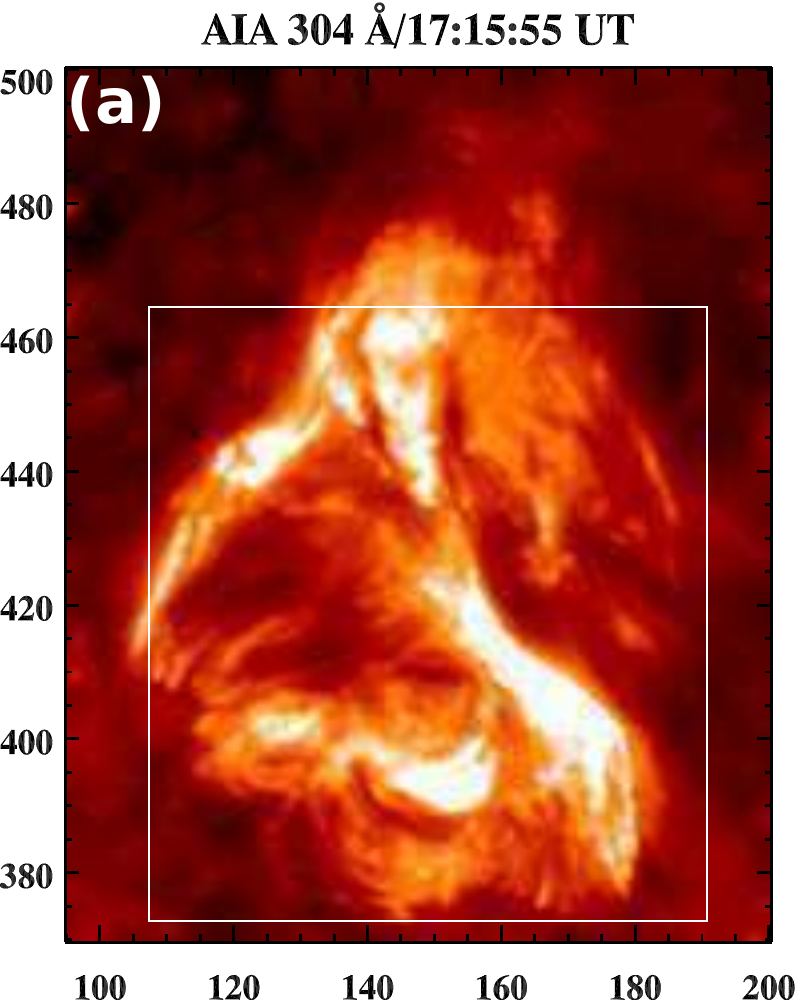}
\includegraphics[width=5.5cm]{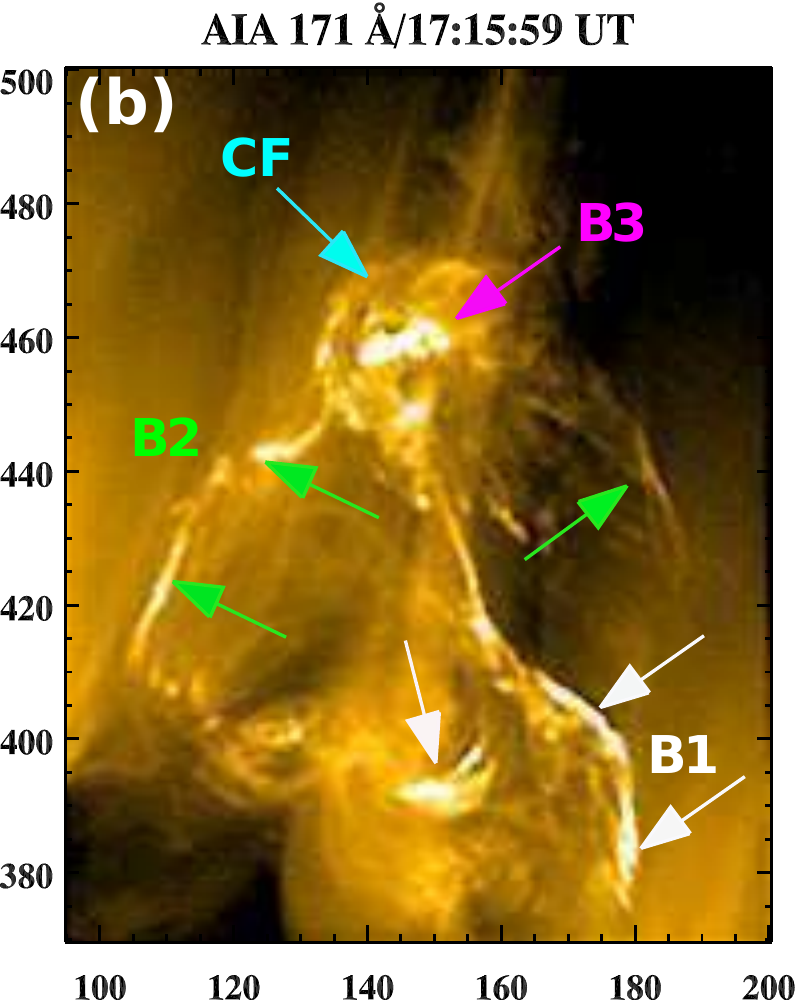}
\includegraphics[width=5.5cm]{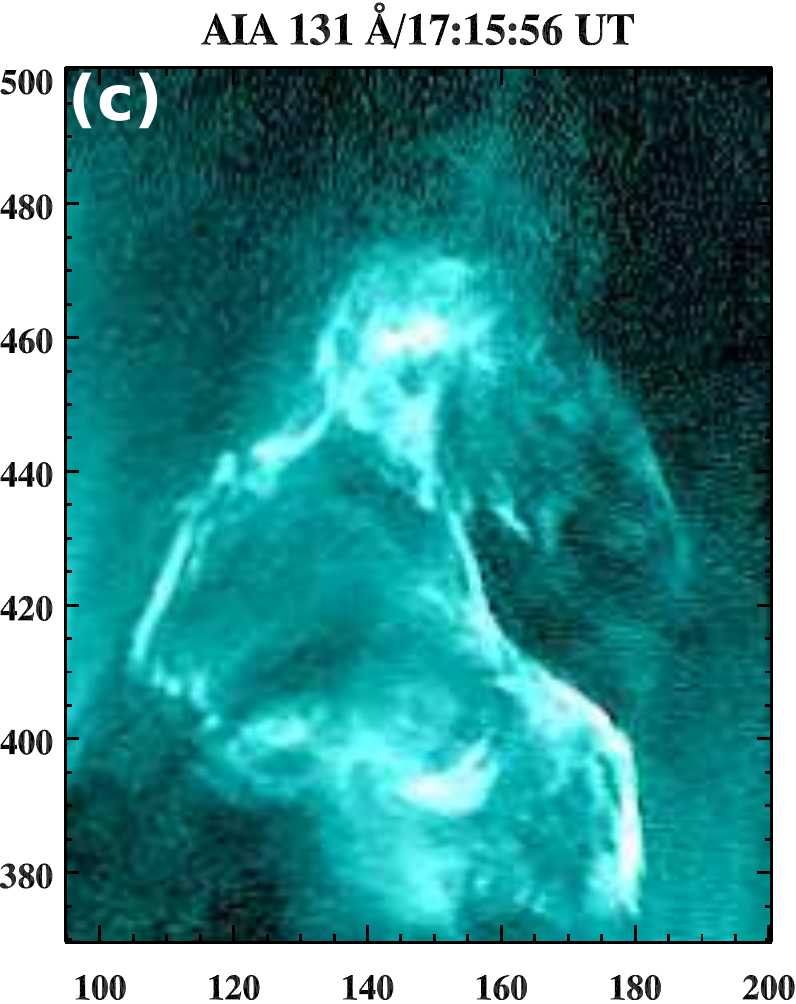}
}
\caption{Selected AIA 304, 171, and 131 \AA~ images prior to jet onset, showing the locations of brightenings B1 (white arrows), B2 (green arrows), and B3 (magenta arrow). The white box in (a) outlines the area used to calculate the integrated intensity profiles shown in Figures \ref{st}(d) and (f). 
} 
\label{bright}
\end{figure*}
\begin{figure*}
\centering{
\includegraphics[width=16cm]{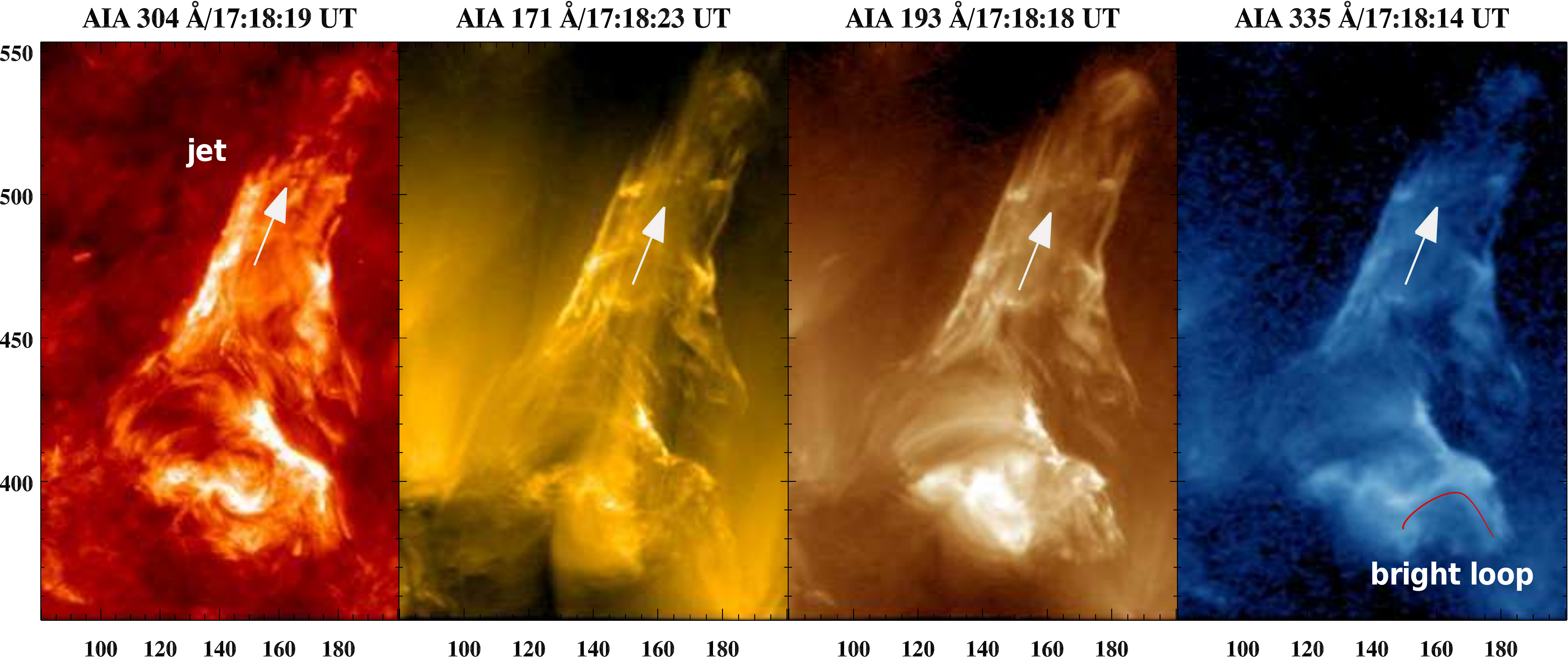}
}
\caption{AIA 304, 171, 193, and 335 \AA~ images showing the jet $\sim$1 min after eruption onset. X and Y axes are labeled in arcsecs. The full dynamic evolution is shown in the movie accompanying Figure \ref{aia1}.
} 
\label{jet}
\end{figure*}
\begin{figure}
\centering{
\includegraphics[width=9cm]{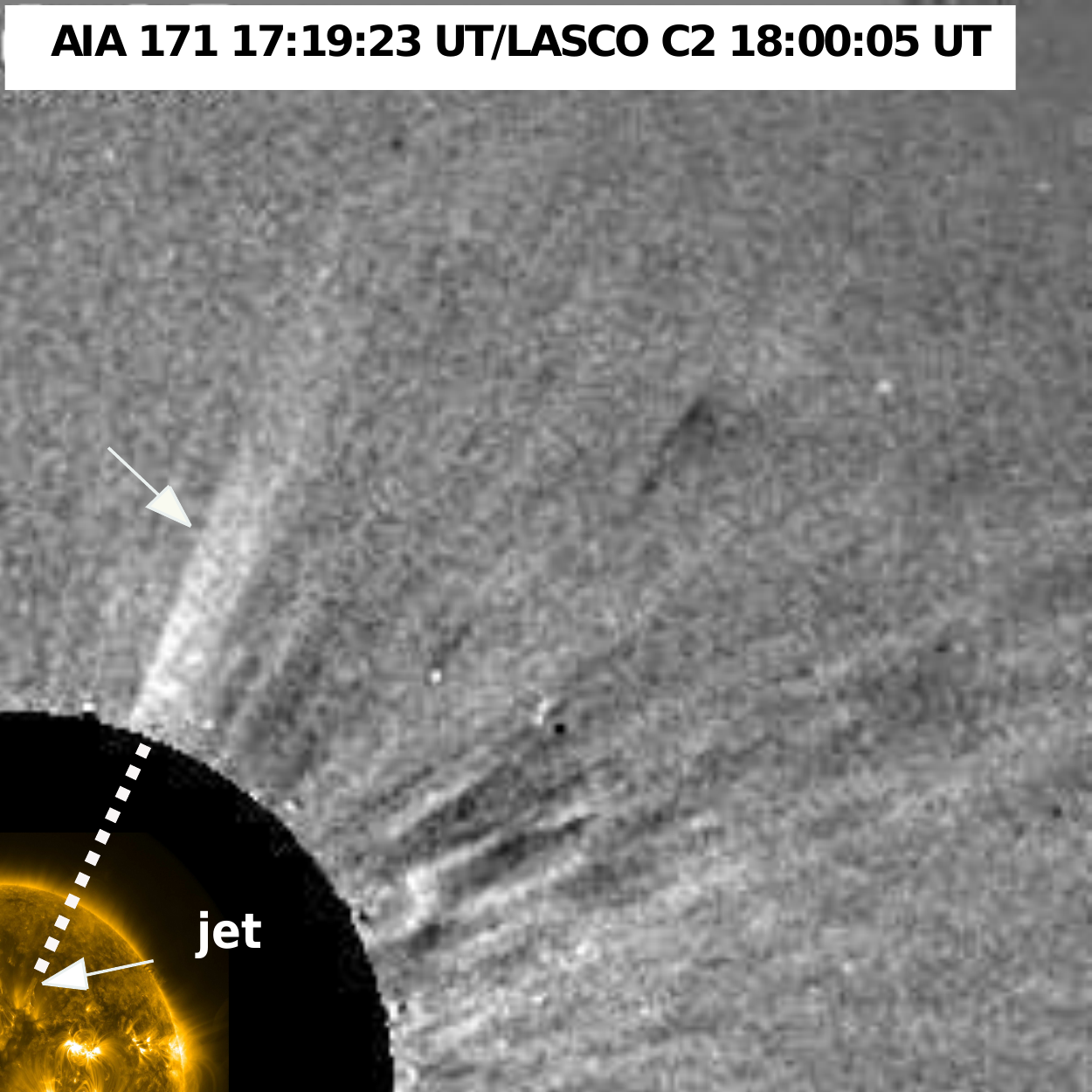}
}
\caption{AIA 171 \AA~ and LASCO C2 coronagraph images showing a narrow CME (marked by arrow) associated with the jet. The dashed line represents a linear extrapolation back to the CME source region, which lines up well with the EUV jet.
} 
\label{cme}
\end{figure}

\begin{figure}
\centering{
\includegraphics[width=8.5cm]{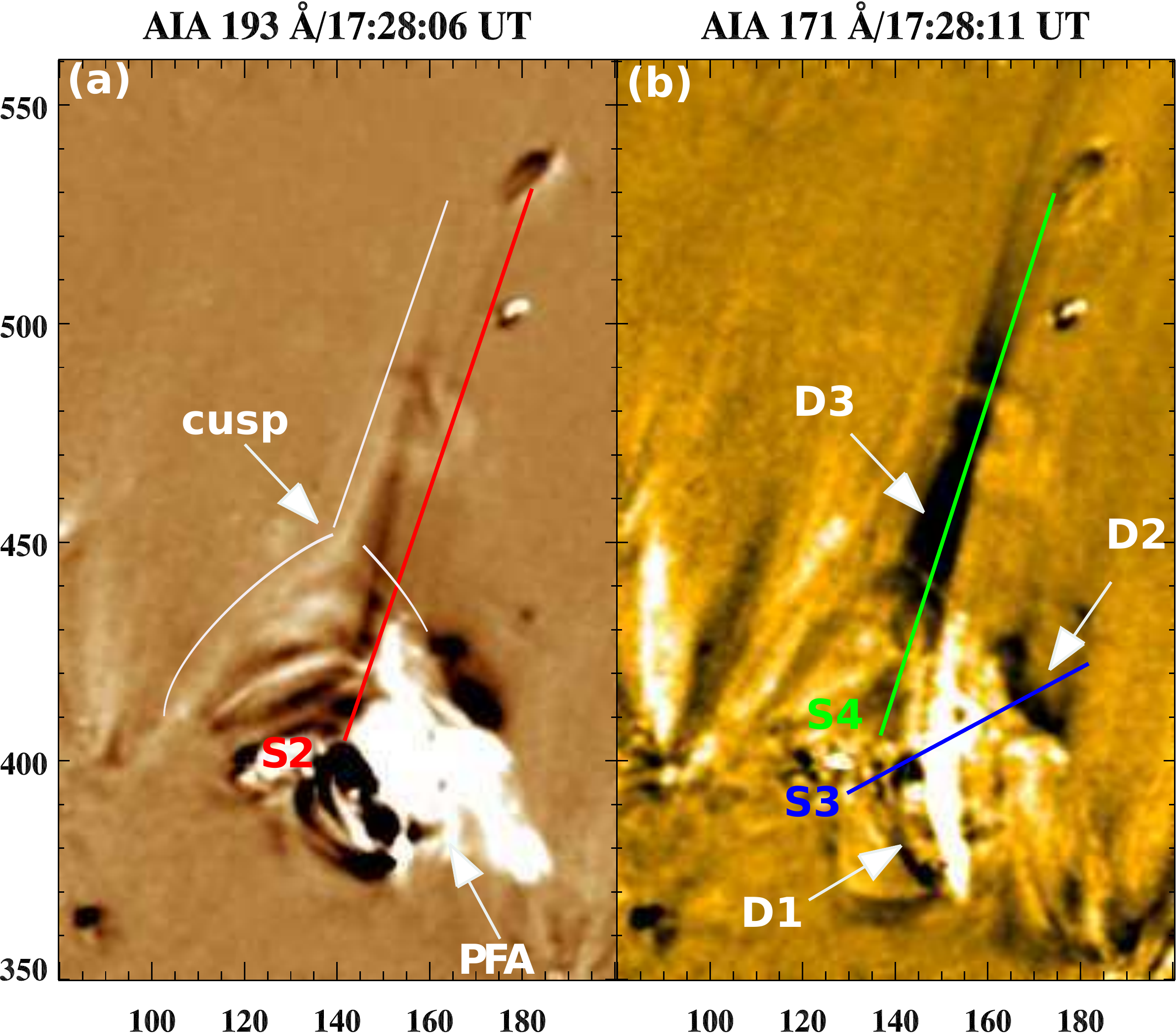}
}
\caption{AIA 193 and 171 \AA~ base-difference images revealing the coronal dimming regions D1, D2, and D3. The base image time is $\sim$16:40 UT. S2 (red), S3 (blue), and S4 (green) are the slices used to create the time-distance intensity plots in Figure \ref{st}(d-f). PFA is the post-flare arcade brightening. X and Y axes are labeled in arcsecs. The full temporal evolution, revealed by base-difference images in the interval 16:40 UT to 17:30 UT, is shown in the accompanying movie. 
} 
\label{dim}
\end{figure}
\begin{figure*}
\centering{
\includegraphics[width=5cm]{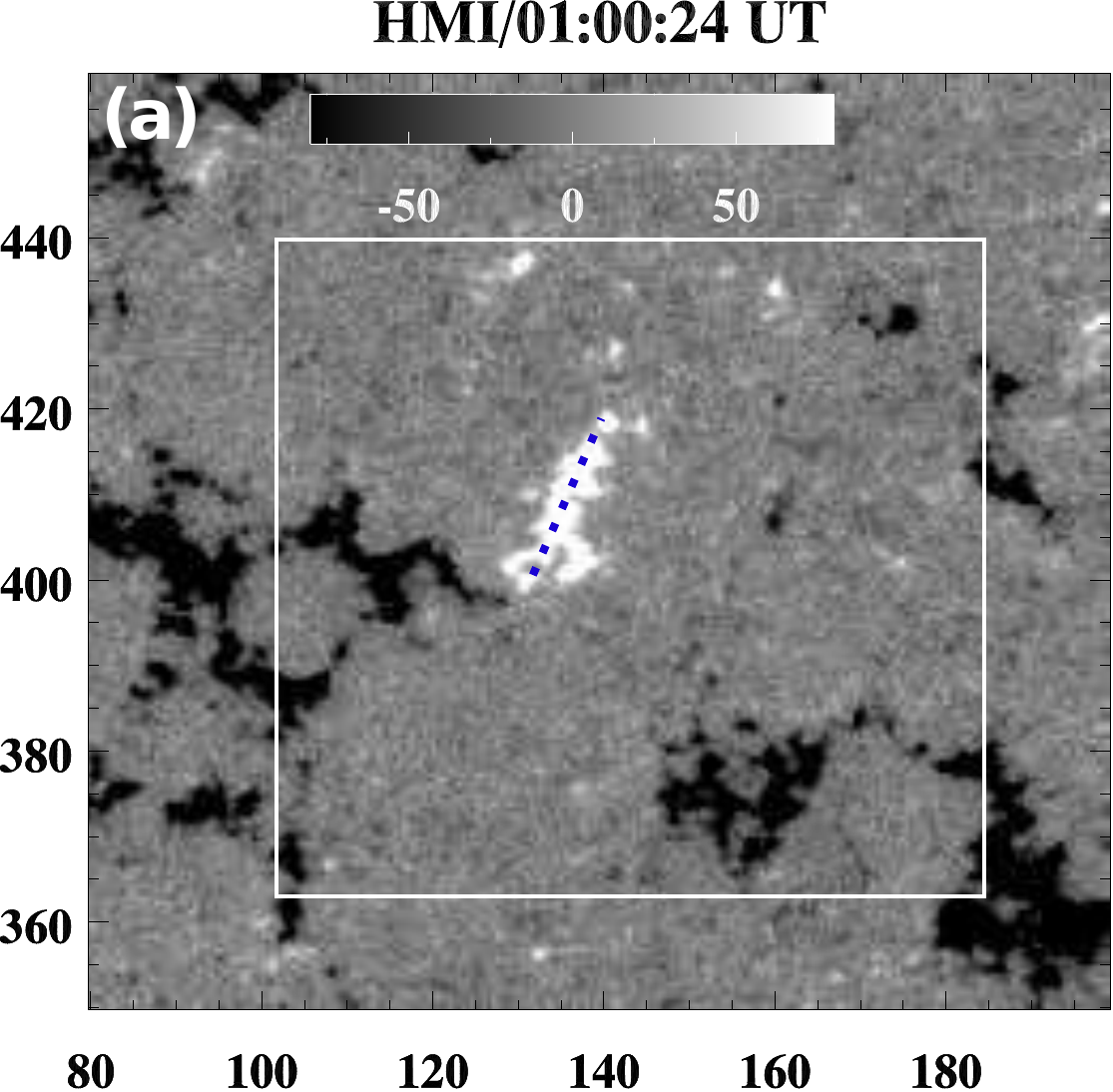}
\includegraphics[width=5cm]{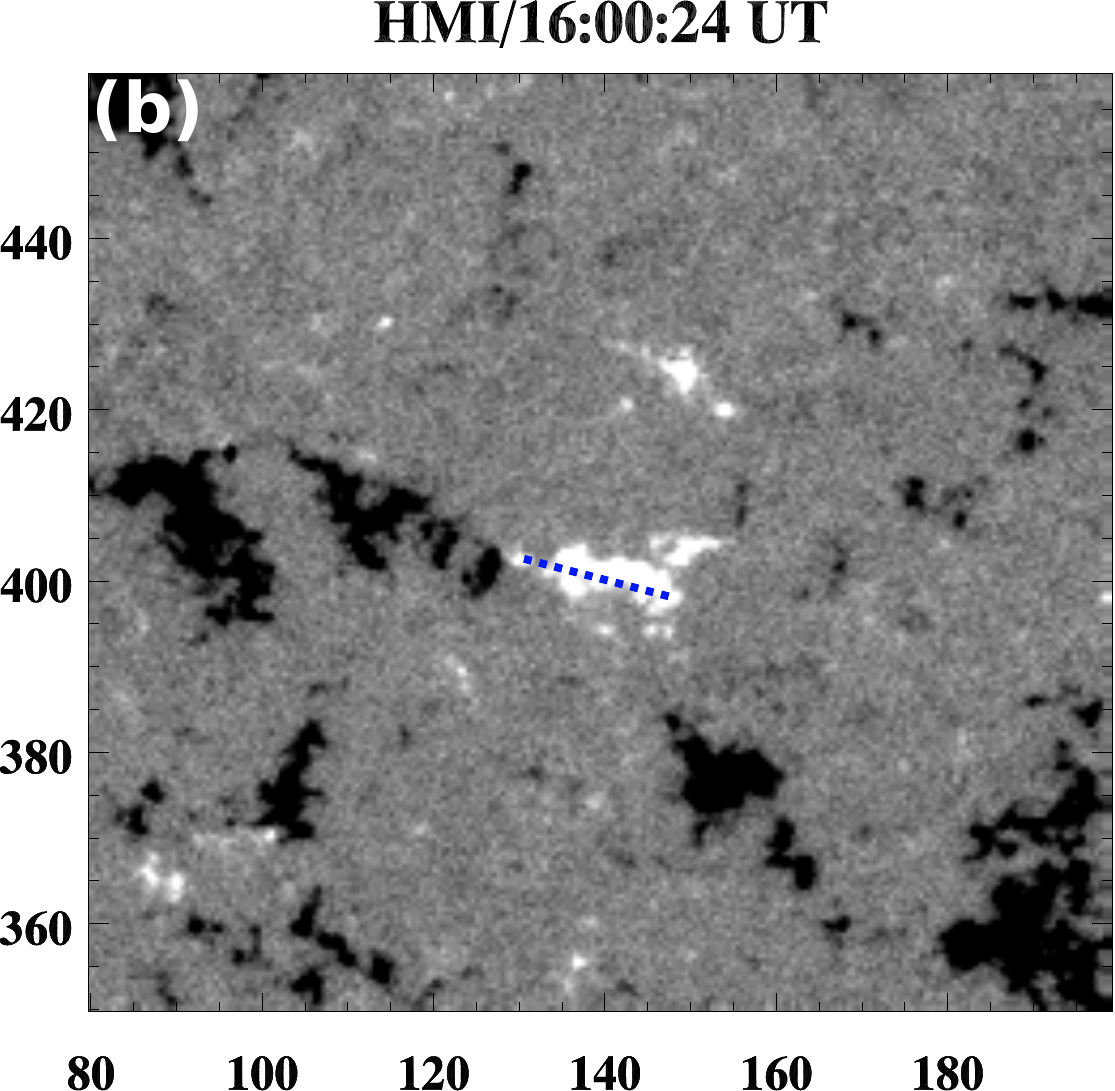}
\includegraphics[width=5cm]{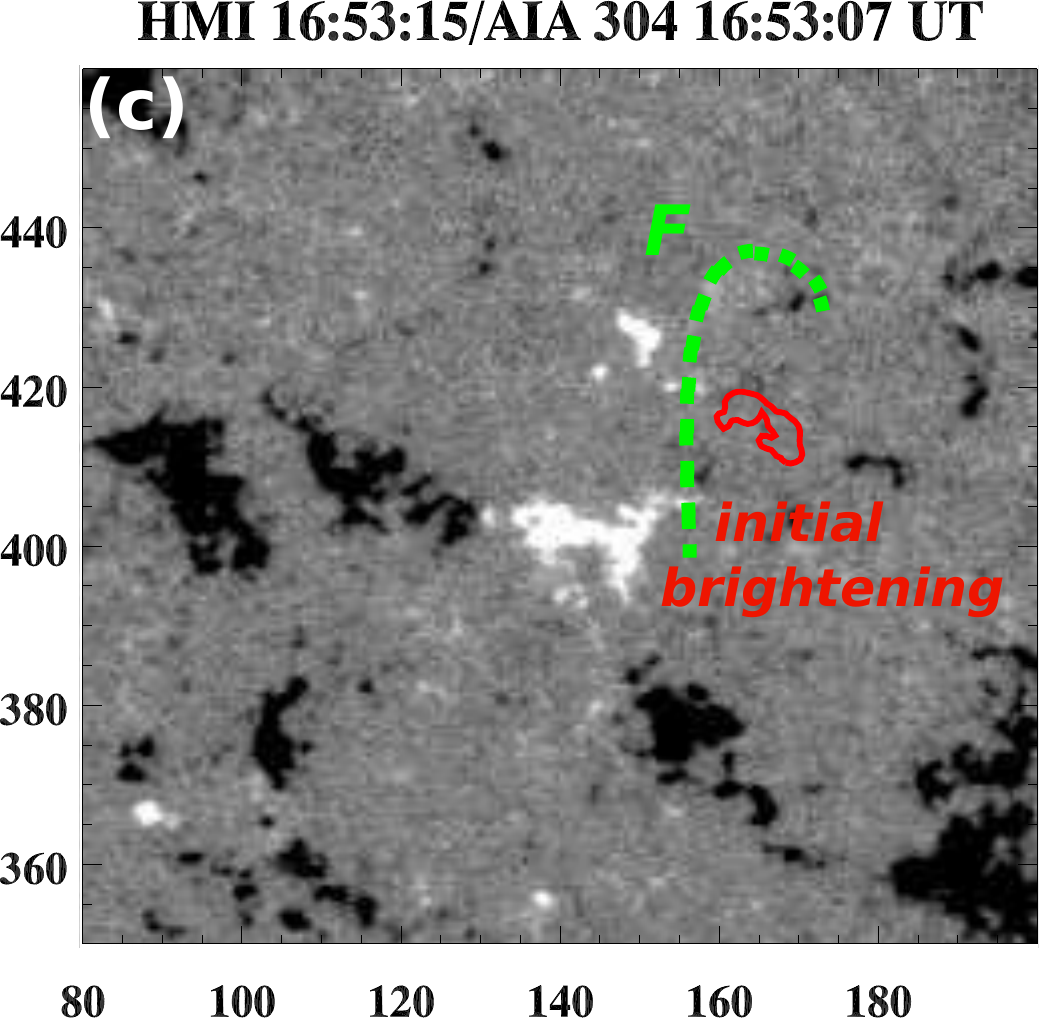}

\includegraphics[width=5cm]{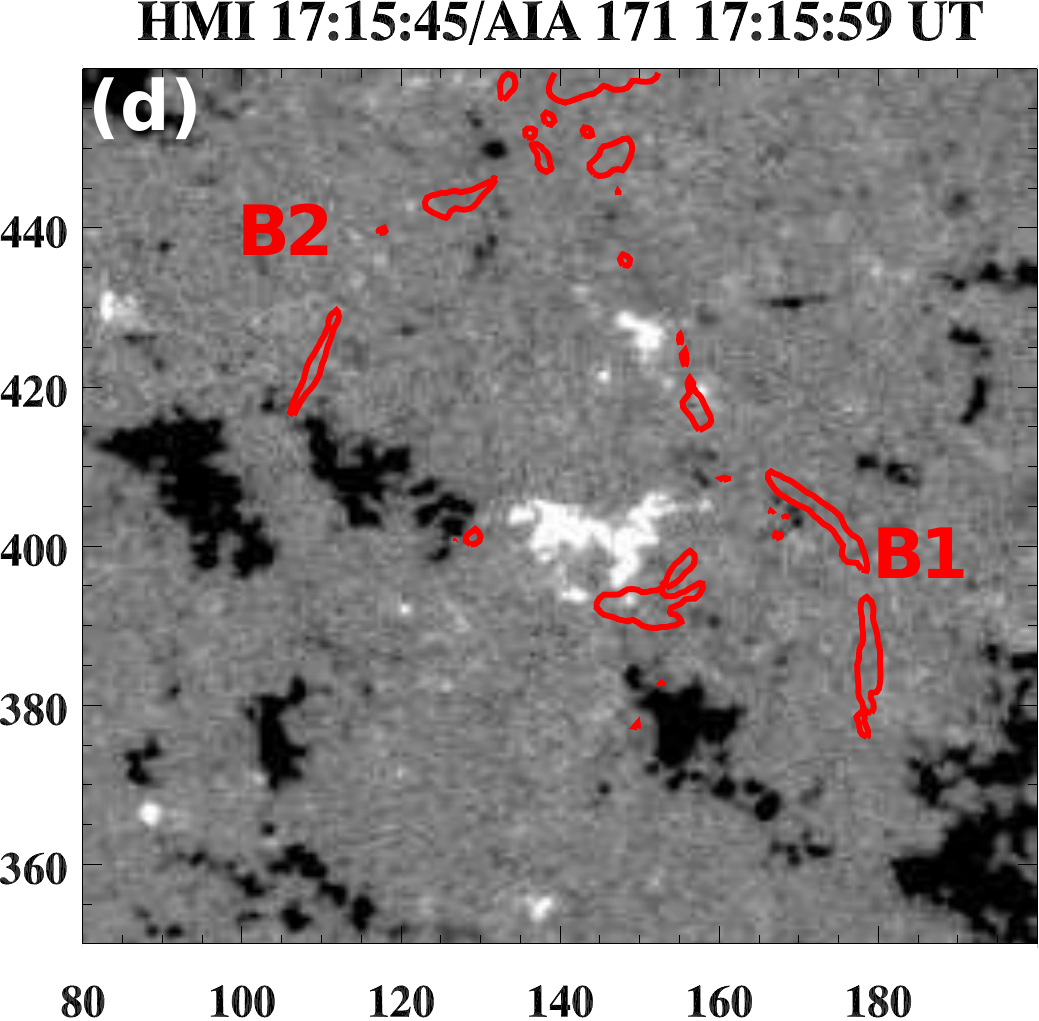}
\includegraphics[width=5cm]{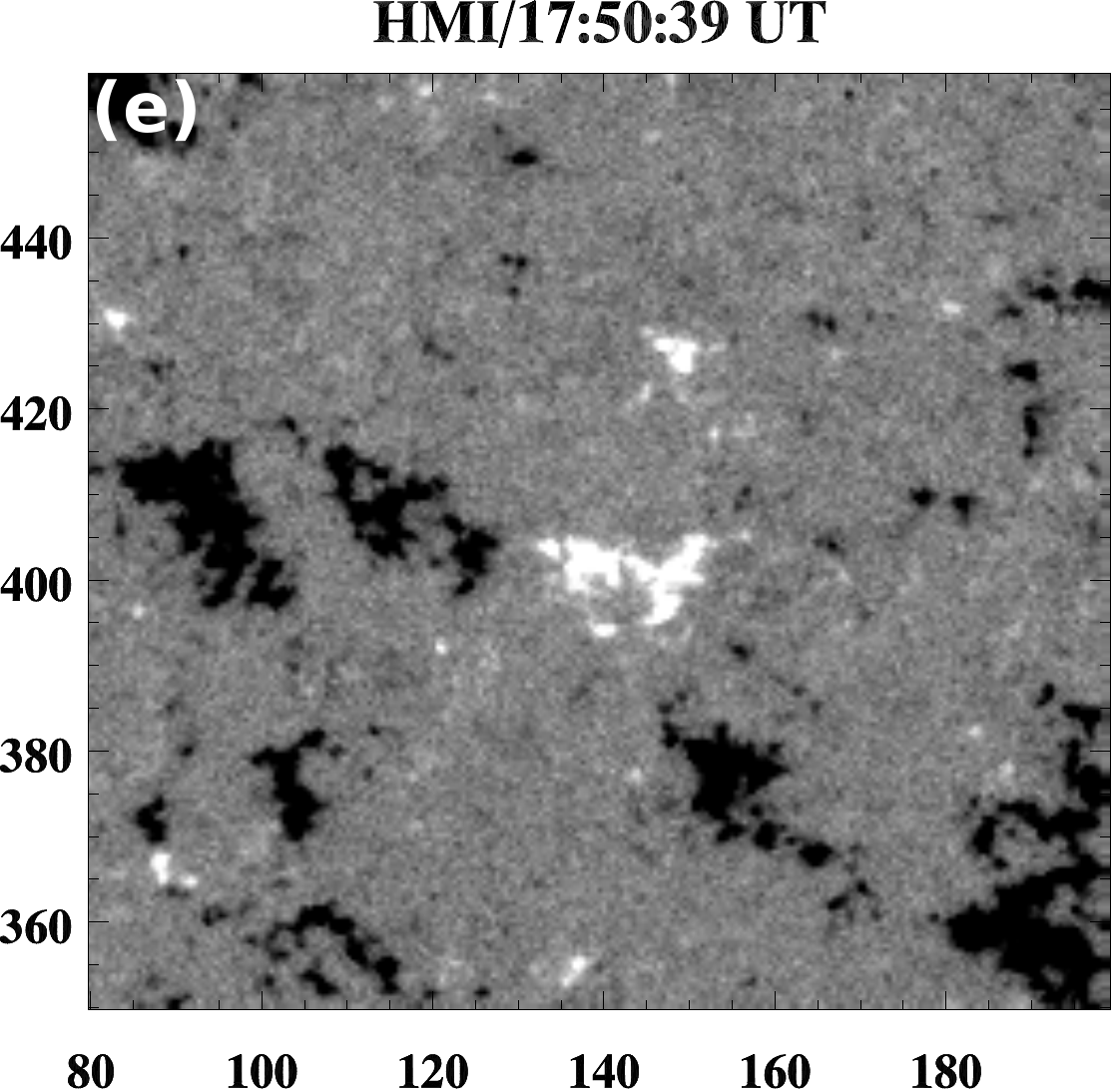}
\includegraphics[width=5cm]{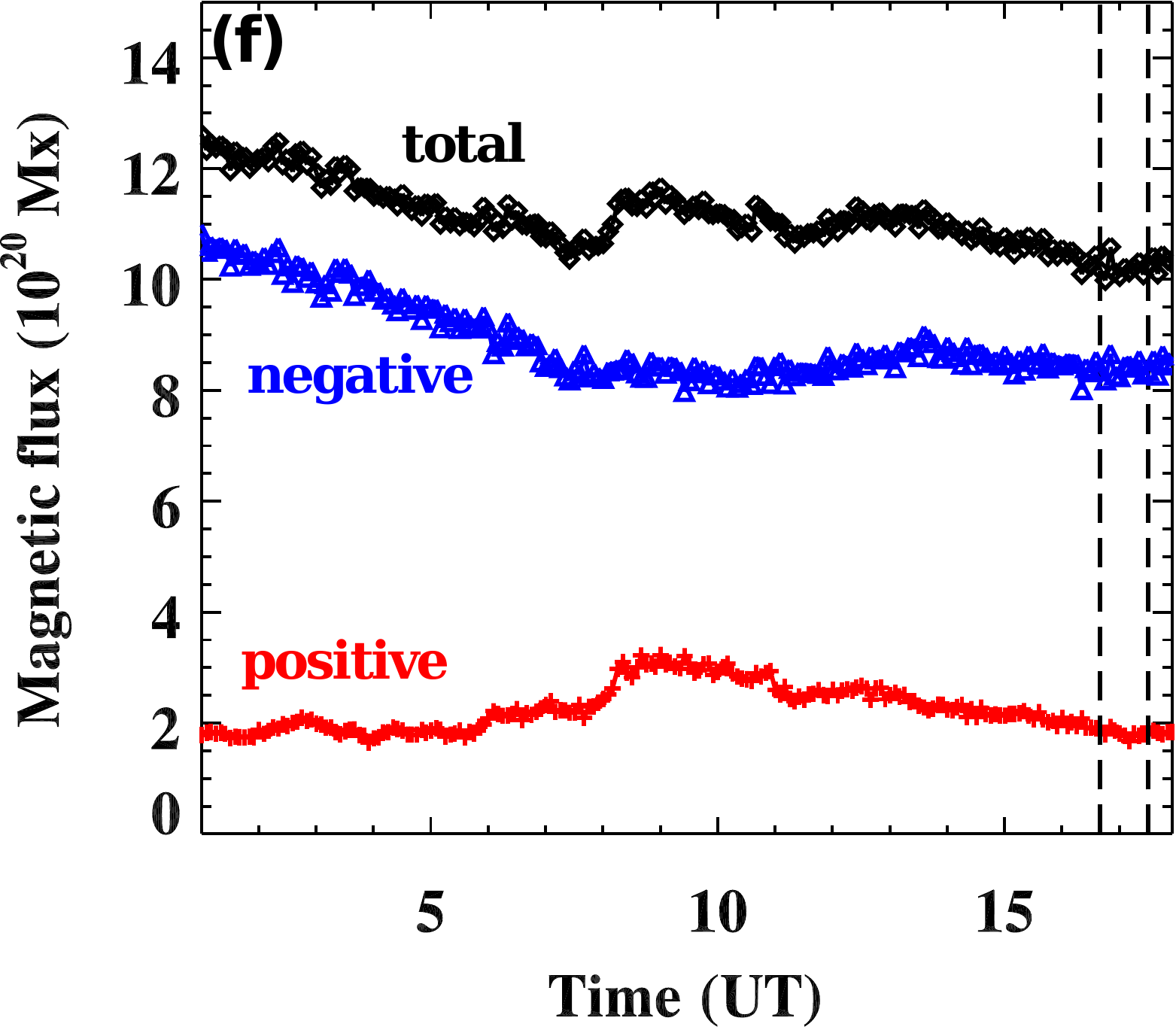}
}
\caption{(a-e) HMI line-of-sight magnetograms of the jet source region before, during, and after the jet activity (01:00-17:50 UT). The red contours over the HMI magnetograms in (c) and (d) mark the outlines of initial brightenings below the filament (F; green dashed line in (c)) and the coronal brightenings (B2) and flare ribbons (B1) associated with the eruption, respectively. The short blue dotted lines in (a) and (b) indicate the main axis of the positive polarity patch, which apparently rotated significantly a few hours before the event.  X and Y axes are labeled in arcsecs. (f) The positive, absolute negative, and total flux profiles (01:00 UT-17:50 UT) close to the eruption site were extracted from within the white rectangular box in (a). The two vertical dashed lines (at 16:40 and 17:30 UT) indicate the duration of the eruption. The full temporal evolution of the line-of-sight magnetic field from 01:32 UT to 18:15 UT is shown in the accompanying movie. 
} 
\label{hmi1}
\end{figure*}
\begin{figure*}
\centering{
\includegraphics[width=5.5cm]{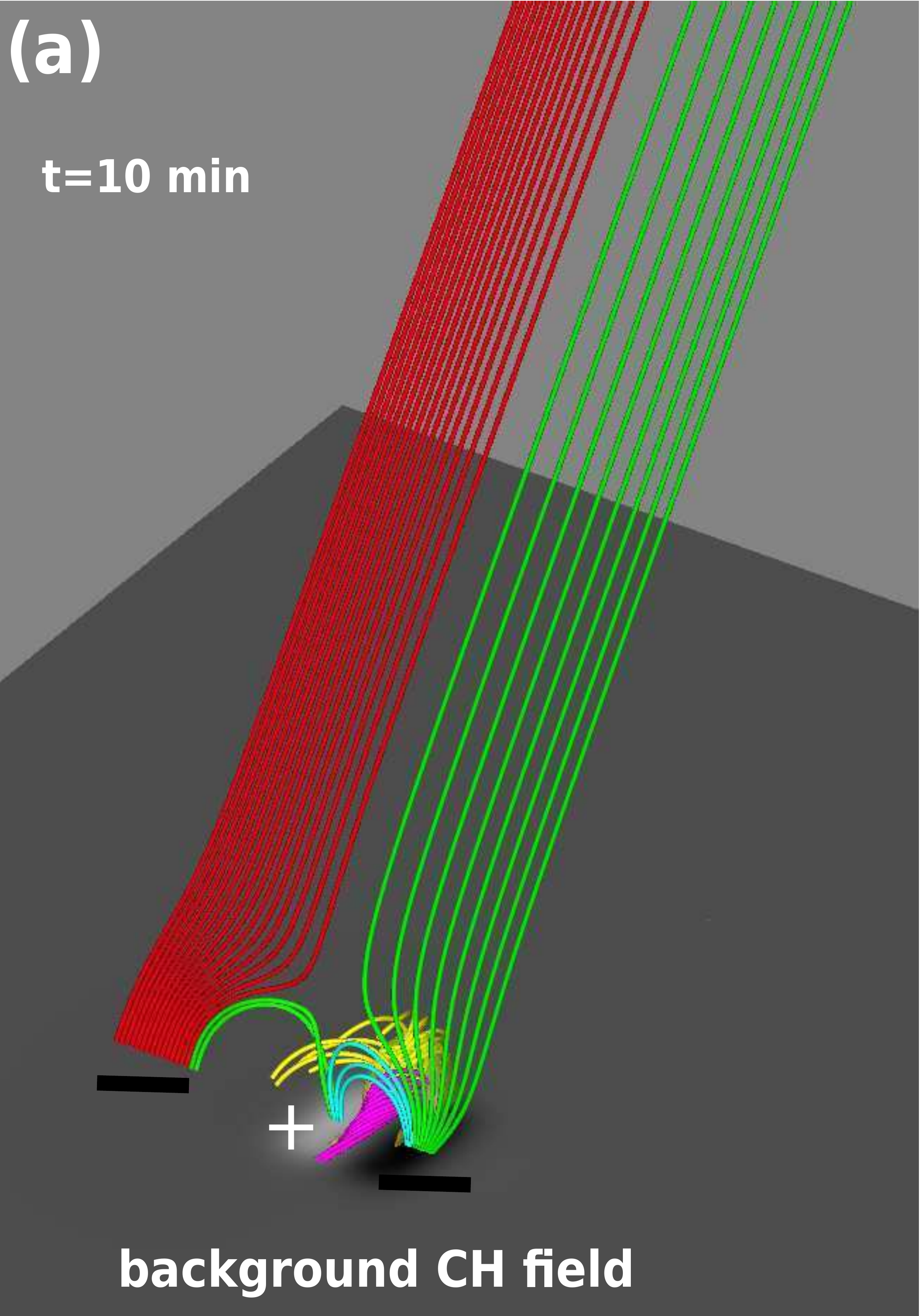}
\includegraphics[width=5.5cm]{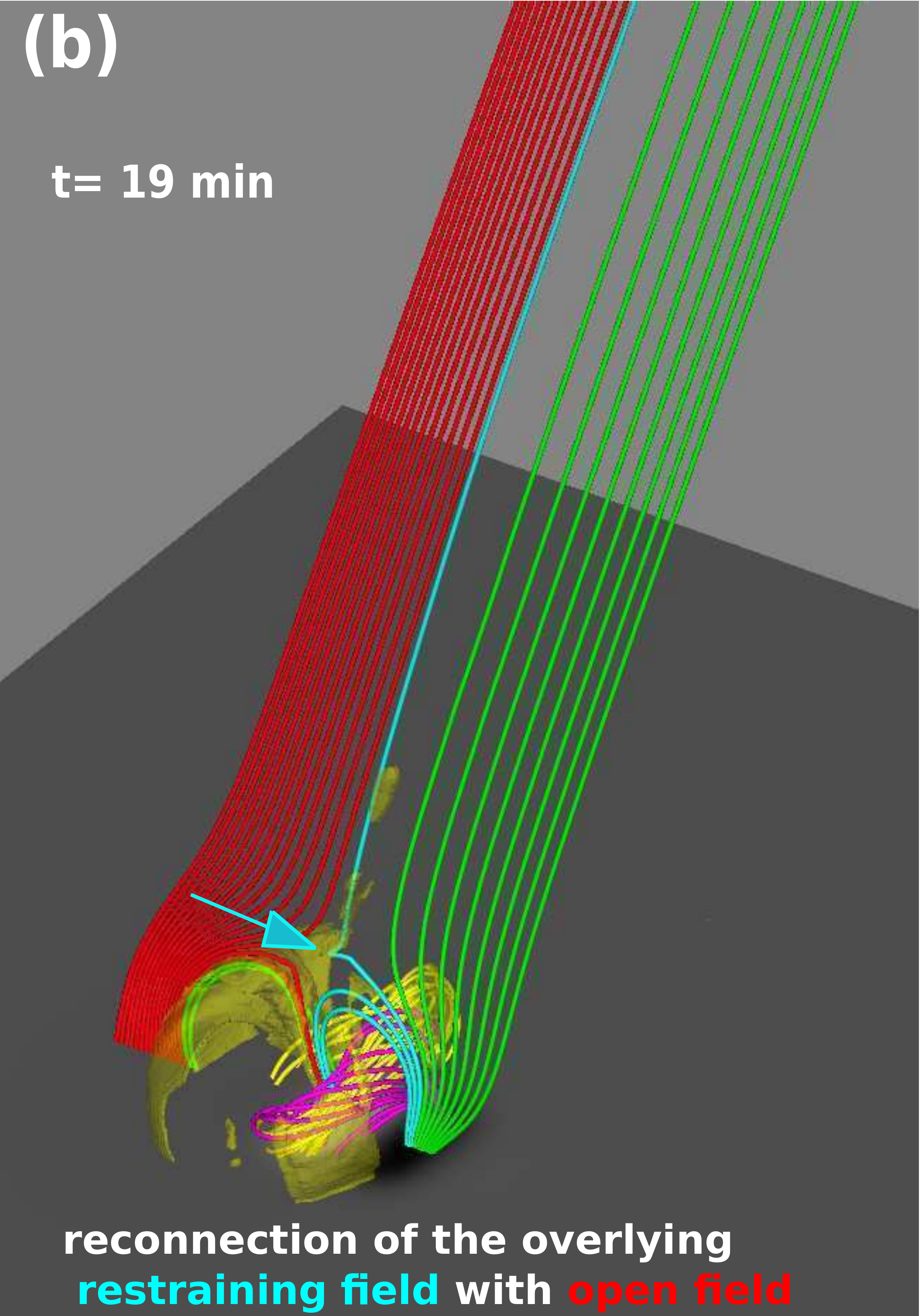}
\includegraphics[width=5.5cm]{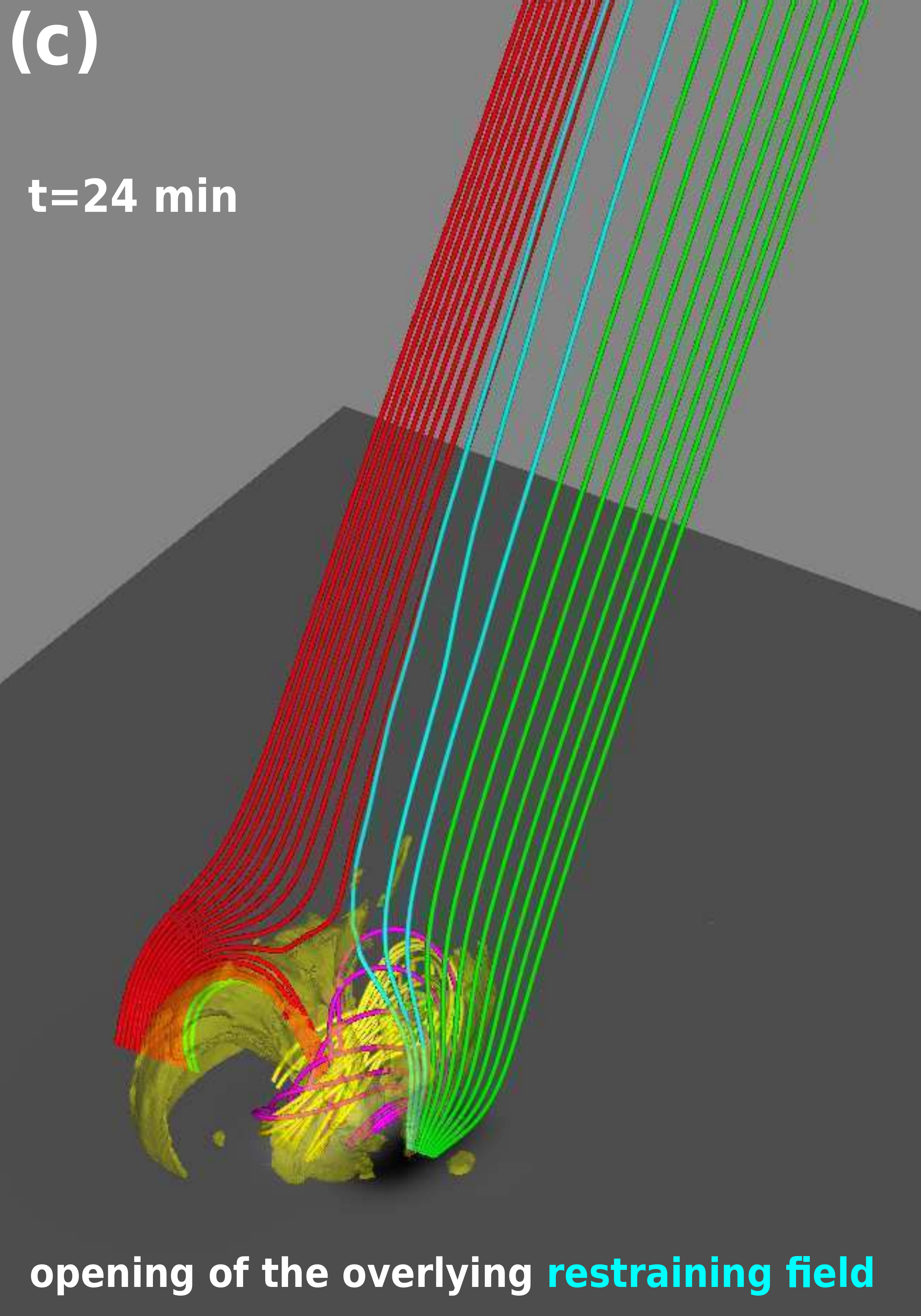}

\includegraphics[width=5.5cm]{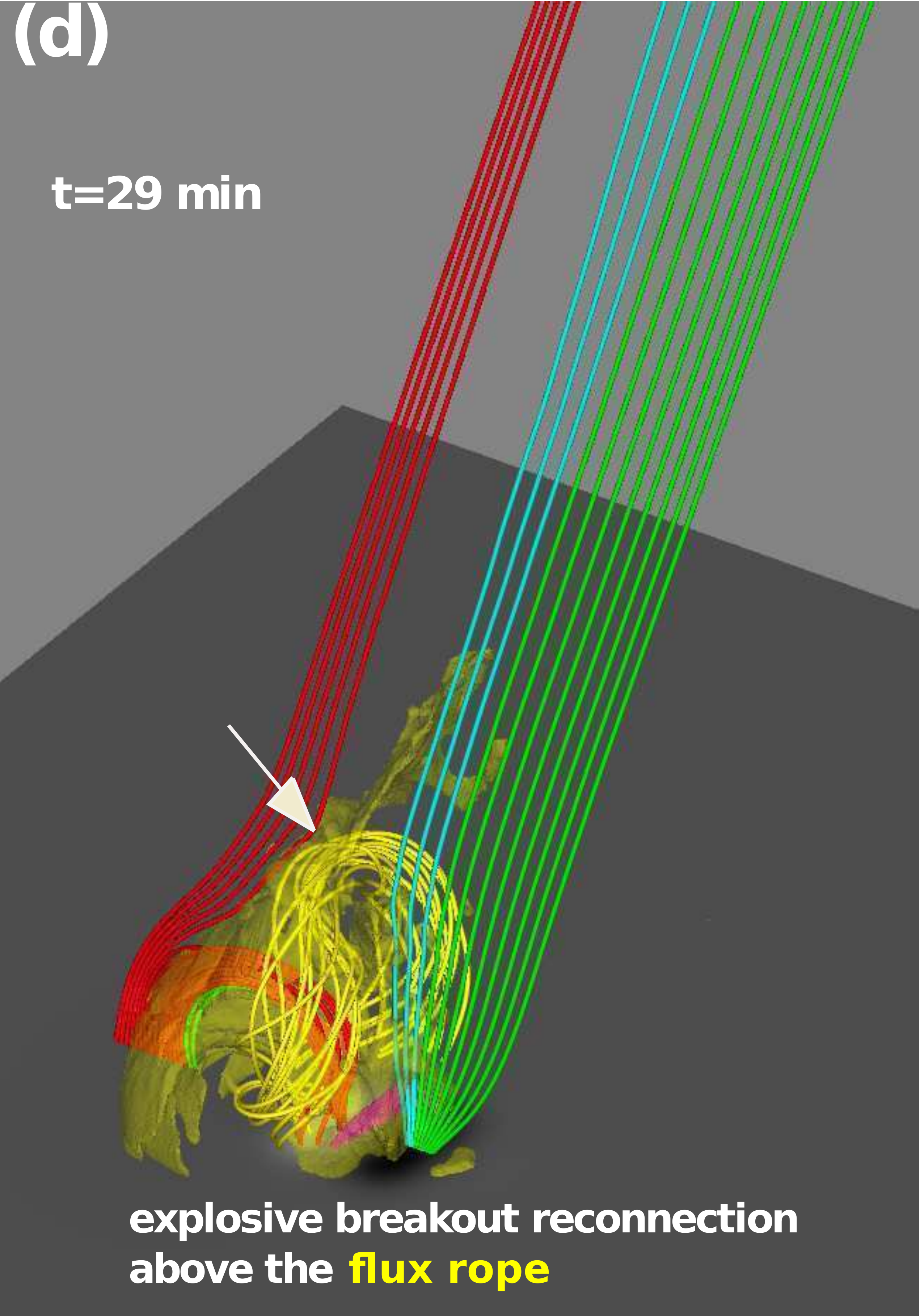}
\includegraphics[width=5.5cm]{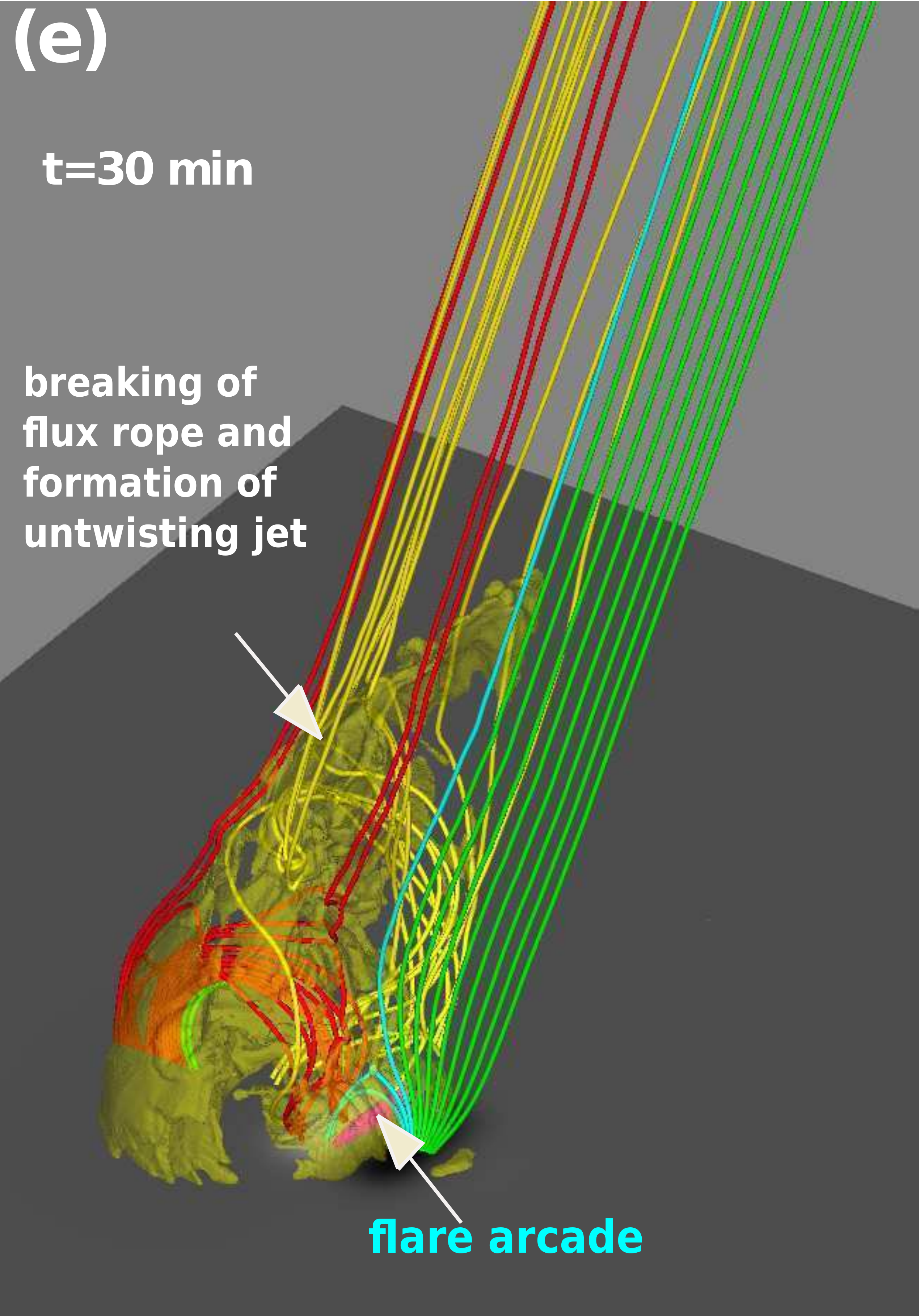}
}
\caption{Selected panels from a 3D MHD simulation of the breakout jet model \citep{wyper2017}. Isosurfaces show enhanced plasma density. (An animation of this Figure is available online from the \citet{wyper2017} paper. The sampling time is 40 s between each frame.)
} 
\label{sim}
\end{figure*}
\begin{figure*}
\centering{
\includegraphics[width=10.5cm]{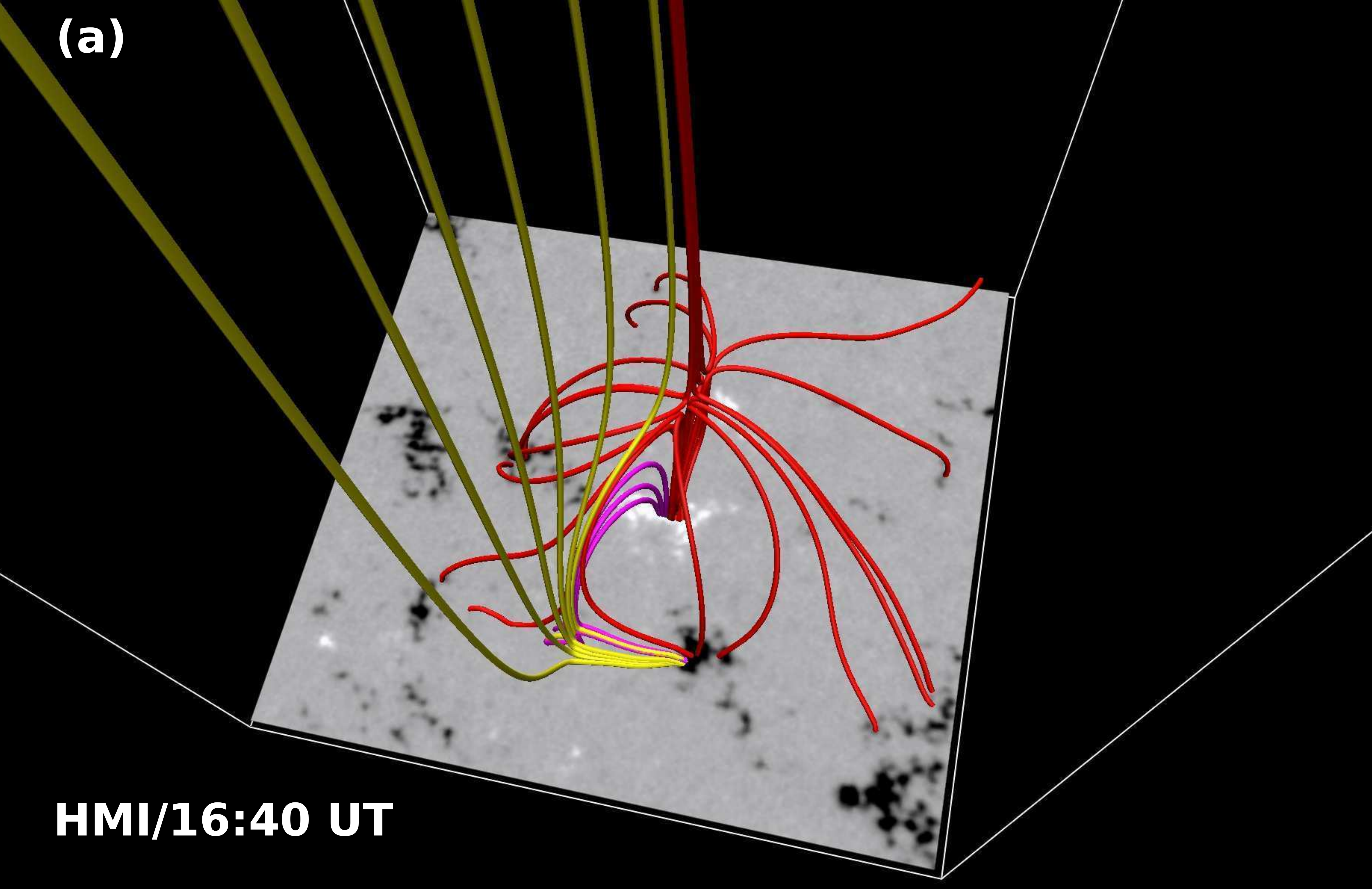}
\includegraphics[width=6.8cm]{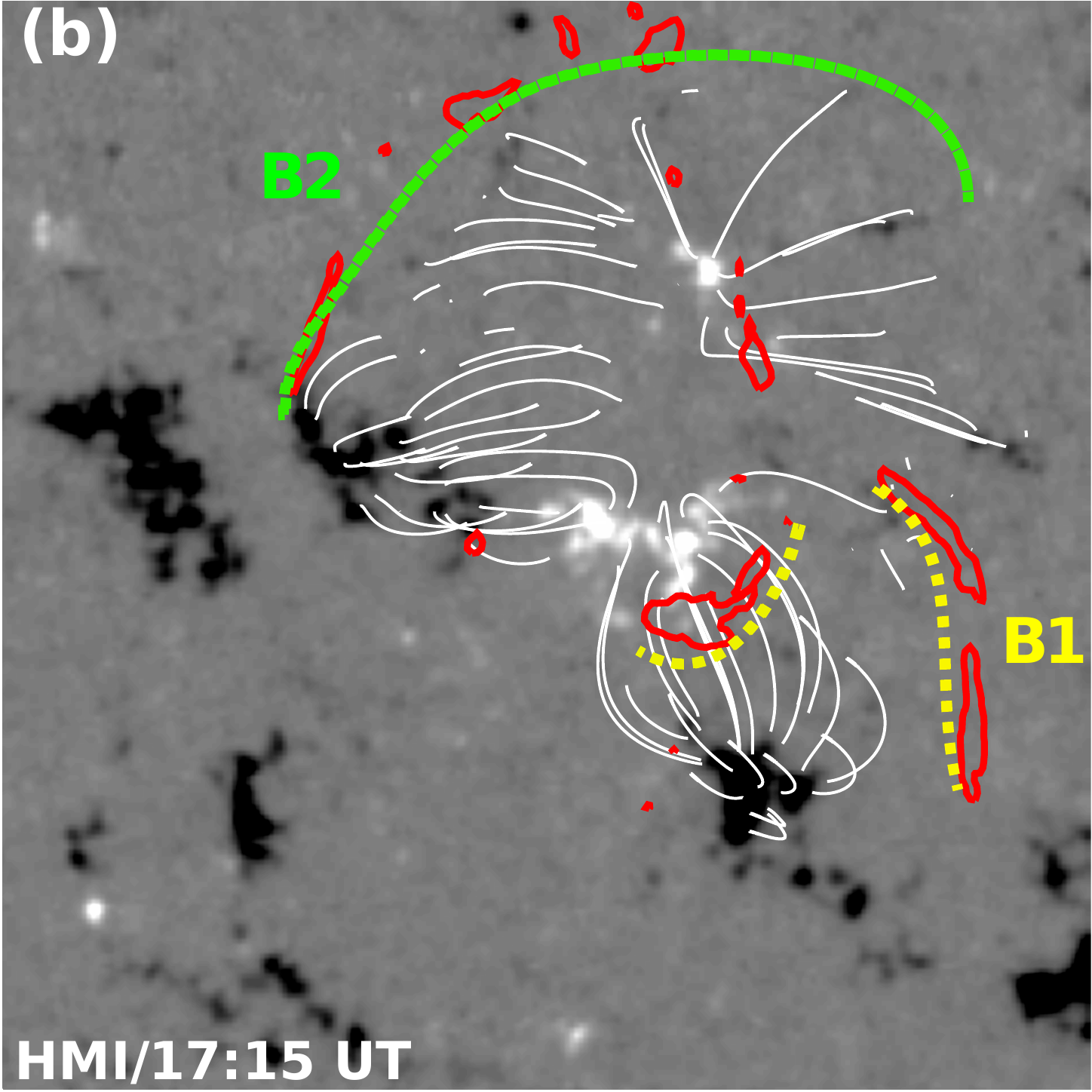}
}
\caption{Selected field lines from potential-field extrapolations of the magnetic field in the eruption site, based on HMI magnetograms at (a) 16:40 UT and (b) 17:15 UT. In (a) the red field lines originate from around the null point above the central positive-polarity patch, the yellow open field lines are drawn from negative polarity concentrations outside the fan surface, and magenta closed field lines are drawn from neighboring negative polarities inside the fan. The yellow and green dashed curves in (b) indicate the locations of brightenings B1 (flare ribbons) and B2 (remote ribbons and fan brightening), respectively. The red contours are the AIA 171 \AA~ brightenings (Fig. \ref{bright}(b)) during the explosive breakout reconnection. 
} 
\label{hmi2}
\end{figure*}

\section{OBSERVATIONS}\label{obs}
We used the {\it Solar Dynamics Observatory} (SDO)/Atmospheric Imaging Assembly (AIA; \citealt{lemen2012}) full-disk images of the Sun (field-of-view $\sim$1.3~R$_\odot$) with a spatial resolution of 1.5$\arcsec$ (0.6$\arcsec$~pixel$^{-1}$) and a cadence of 12~s. We utilized AIA 304~\AA\ (\ion{He}{2}, at temperature $T\approx 0.05$~MK), 171~\AA\ (\ion{Fe}{9}, $T\approx 0.7$~MK), 211~\AA\ (\ion{Fe}{14}, $T\approx 2$~MK), 335~\AA\ (\ion{Fe}{16}, $T\approx 2.5$~MK), 94~\AA\ (\ion{Fe}{10}, \ion{Fe}{18}, $T\approx 1$ and 6.3~MK, respectively), 131~\AA\ (\ion{Fe}{8}, \ion{Fe}{21},  \ion{Fe}{23}, $T\approx 0.4$, 10, 16~MK, respectively), and 193~\AA\ (\ion{Fe}{12}, \ion{Fe}{24}, $T\approx  1.2$ and $\approx 20$~MK, respectively) images. A new 3D noise-gating technique \citep{deforest2017} was used to clean the AIA images and the Helioseismic and Magnetic Imager (HMI; \citealt{schou2012}) magnetograms, which were analyzed at a 45-s cadence.
 
\subsection{AIA observations}
A big equatorial coronal hole extended from near disk center to the north pole on 9 January 2014. Within the dark CH, shown in Figure \ref{aia211}(a), the jet source region is marked by a red rectangular box. Figure \ref{aia211}(b) shows an enlarged view of the jet source region, with overlaid HMI magnetogram contours ($\pm$50 Gauss) of positive (blue) and negative (green) polarities to characterize the line-of-sight photospheric magnetic field in this region at approximately the same time as the AIA image. The background magnetic polarity of the CH is negative. The local magnetic configuration consists of at least one compact positive-polarity region (+) surrounded by many small negative-polarity regions (-): a classic embedded bipole \citep{antiochos1996}.

We used EUV images in different wavelengths to infer the evolution of the magnetic structures at the eruption site from the chromosphere to the corona, prior to and during the jet.  Figure \ref{aia1} and the accompanying movie show AIA images in 304, 171, 193, and 335 \AA~ channels, in a sequence of increasing temperature response from left to right and increasing time from top to bottom. The earliest 304 \AA~ panel shows the pre-eruption configuration of the jet source region at $\sim$16:40 UT. Note the dark ``mini-filament" in absorption (F, marked by an arrow and a dashed outline) lying along the polarity inversion line (PIL). Inspection of earlier AIA data reveals that the mini-filament first became visible $\sim$19 hours before the eruption. The co-temporal images at coronal temperatures (171, 193, and 335 \AA) exhibit a bright, S-shaped, sigmoidal structure (white dashed line in the earliest 193 \AA~ panel) in which the small filament was embedded.  In addition, these images and the 211 \AA~ image (Fig. \ref{aia211}(b)) reveal a bright, quasi-vertical linear feature (``spire", yellow dashed line in the earliest 193 \AA~ panel) above a dome-shaped structure (white arc in the earliest 193 \AA~ panel). Within the dome resided a brighter set of long-lived, low-lying loops connecting from the central positive-polarity to the surrounding negative-polarity concentrations. The middle panels of Figure \ref{aia1} show brightenings below the filament at $\sim$16:53 UT, seen most clearly in the accompanying movie. The dark filament rose slowly until $\sim$17:11 UT, trailed by a lengthening, bright, linear feature, while the rising bright structure (dotted green arc) surrounding the filament became circular.  

Figure \ref{blobs} and the accompanying movie display the AIA 171, 193, and 131 \AA\ unsharp-masked images during 17:12:47 -17:13:59 UT. Multiple bright blobs (marked by arrows) appeared in all AIA channels below the circular feature (CF) surrounding the dark mini-filament (clearest in the AIA 193 and 131 \AA~ channels), simultaneous with the formation of extended, narrow brightenings at the surface (see Figure 2, bottom panels). The $\sim$2-3 arcsec-wide blobs propagated upward and downward along the bright, increasingly extended, inverted-V shaped structure below the CF, which is marked by the green arrow in the bottom panel of Figure \ref{blobs}. Upward-moving blobs are visible in this region until $\sim$17:16 UT, after the fast rise of the CF began but before jet onset. 

After the leading edge of the circular feature touched the overlying structures near the long spire at $\sim$17:12 UT, a phase of explosive activity began. The low-lying loops and the outline of the CF brightened substantially (see Figure \ref{blobs}), and the CF rose more rapidly.  During this interval we also observed significant leftward deflection of the bright spire from its initial location, as shown in the 171 \AA~ running-difference images of Figure \ref{def}(a,b). By $\sim$17:16 UT, the CF reached its maximum height, and multiple transient brightenings appeared simultaneously in the 304, 171 and 131 \AA~ images (Fig. \ref{bright}). Two bright arcs labeled B1 (white arrows) appeared below the thin linear feature that hosted the traveling blobs; a thin, discontinuous, curved arc labeled B2 (green arrows) appeared on both sides of the CF; and a compact bright feature labeled B3 appeared near the apparent intersection between the CF and the spire. 

Figure \ref{jet} shows the jet (direction indicated by an arrow) in different AIA channels at 17:18 UT; the temporal evolution is most evident in the accompanying movie. These images reveal a wide jet extending from a bright core, which consists of the aforementioned inverted-V shaped structure above 2 bright bands joined by a hotter (335 \AA~ emitting) loop or arcade. The overall spatial distribution of emission is similar in all channels, but the relative brightness of specific features varies, indicating that plasma at different temperatures and/or densities co-existed in different locations. The apparent rotation is from rear to front as it progresses from left to right, i.e., counterclockwise or right-handed, as seen in the movie accompanying Figure \ref{aia1}.

A narrow CME was associated with the jet and detected by the LASCO C2 coronagraph (2-6 R$_\odot$; \citealt{brueckner1995}); an AIA 171 \AA~ (17:19:23 UT) and LASCO C2 white-light (18:00:05 UT) composite image is shown in Figure \ref{cme}. The CME direction is marked by a dashed line projected from the narrow jet in the C2 field of view back to the AIA field of view; note that this line intersects the jet source region. The leading edge of the jet reached at least $\sim$4-5 R$_\odot$ in C2. 

Figure \ref{dim} shows AIA 171 and 193 \AA~ base-difference images after the jet at $\sim$17:28 UT.  The associated movie reveals the formation of strong dimming regions near the spire (D3) and at the ends of the initial sigmoid (D1, D2). To better understand the origin of these dimmings and displacements during the dynamic event, we created time-distance intensity maps along slices S2, S3, and S4, which are discussed in \S \ref{results}. 
\subsection{HMI observations}
To investigate whether flux emergence, cancellation, or footpoint motions played a role in triggering the mini-filament eruption, we analyzed HMI magnetograms during 01:00 UT to 17:50 UT. During the interval before $\sim$16:00 UT, the HMI movie accompanying Figure \ref{hmi1} shows that the elongated central positive-polarity patch changed from a north-south orientation to east-west. However, it is unclear whether this reflects actual clockwise rotation or reshuffling and convergence of many smaller flux tubes with like polarity. Figure \ref{hmi1} shows selected magnetograms before, during, and after the eruption (01:00-17:50 UT). The red contours over panels (c),(d) outline the brightenings at the start ($\sim$16:53 UT) and peak ($\sim$17:15 UT) of the jet. Although small concentrations of positive and negative flux evolved constantly, no bipolar concentrations on the scale of the source region appeared or disappeared before or at the time of the eruption. For the area within the white rectangular box in Figure \ref{hmi1}(a), we extracted the positive, absolute negative, and total fluxes during 01:00 UT-17:50 UT (Figure \ref{hmi1}(f)).  Flux emergence or cancellation should increase or decrease both polarities equally and simultaneously, which is not observed.  Therefore, we conclude that the magnetograms do not exhibit any significant large-scale flux emergence or cancellation at the eruption site during the 16 hours leading up to the eruption, as evidenced in Figure \ref{hmi1}(f). 

\section{RESISTIVE-KINK AND BREAKOUT JET MODELS}\label{model}

Based on previous theoretical studies \citep{antiochos1990,antiochos1996}, we developed and advanced the embedded-bipole model for coronal-hole jets \citep{pariat2009,pariat2010,pariat2015,pariat2016,wyper2016a,wyper2016b,karpen2017}. The source region in the model consists of a small, relatively strong concentration of minority-polarity flux embedded in a broad sea of more diffuse majority-polarity flux: the classic 3D fan-spine topology.  The separatrix between the closed bipolar flux system and the surrounding open flux forms a dome-shaped structure with a null point on its surface. Electric current sheets develop readily at the null and separatrix surface through relative displacements of the field inside and outside the dome \citep{antiochos1990,antiochos1996,lau1990}. When these current sheets become sufficiently thin, magnetic reconnection occurs, accompanied by mass motions and plasma heating. In most of the studies listed above, the free energy that drives the jet is provided by rotational footpoint motions over a broad region inside the PIL. Therefore, no filament channel is formed in this model. The twisted flux expands, enlarging the dome and pushing the null higher in the corona. Slow reconnection through the current patch at the null slowly removes the restraining field and drives weak outflows along the spine.  Explosive reconnection occurs only after the twisted closed flux undergoes an impulsive, kink-like instability, forcing the twisted flux against the separatrix.  The resulting reconnection site is driven around the dome as the core untwists, yielding a nonlinear helical Alfv\'en wave that propagates along the reconnected open field lines accompanied by slower upflows of dense plasma. In this {\it resistive-kink jet}, reconnection and ideal instability work together to release the stored energy explosively. 

The {\it breakout jet} model \citep{wyper2017,wyper2018} is a natural extension of the breakout mechanism originally applied to large-scale solar eruptions \citep{antiochos1998,antiochos1999}. In the jet scenario, illustrated by the simulation snapshots in Figure \ref{sim}, the initial configuration is an embedded bipolar region with strong concentrations of both minority- and majority-polarity flux in a background coronal hole, with the usual fan-spine topology and a coronal null.  In the simulation, magnetic shear is added through rotational footpoint motions in a narrow zone at the PIL, causing the overlying restraining field (cyan lines) to expand and create a breakout current sheet at the null. The strongly sheared filament-channel magnetic field (yellow lines, Figure \ref{sim}(b)) is the structure needed to support a filament beneath the overlying restraining field (magenta and cyan lines). As in the resistive-kink jet model, reconnection at this breakout sheet then slowly removes the restraining field and drives slow, narrow plasma outflows (Fig. \ref{sim}(c)). Feedback between the removal of the restraining field and the upward expansion of the sheared field accelerates both processes. As the filament-channel field expands, a current sheet forms beneath it where slow reconnection converts the rising sheared arcade into a flux rope. Only the section of the filament channel with the strongest shear is converted into a rising flux rope, so residual shear remains both beneath the flux rope and, more weakly, in the remainder of the channel. Explosive reconnection occurs only when the flux rope collides with the external open field (red lines) as it reaches the breakout current sheet (Fig. \ref{sim}(d)), releasing a nonlinear Alfv\'en wave and an untwisting Alfv\'enic jet (Fig. \ref{sim}(e)) similar to that seen in our resistive-kink jet model. A mini-flare arcade forms beneath the erupting flux rope, directly analogous to the flare arcade predicted by most CME/eruptive flare models \citep[e.g.,][]{karpen2012}. As the twist is released onto open field lines, the dominant reconnection site and the spine moves along the fan surface from the cusp at the top to the flare current sheet, thus positioning the system to relax back to another equilibrium state. 

The key observable differences between these models are: (1) the breakout jet contains a mini-filament (or at least a filament channel); (2) the breakout jet is accompanied by a flare arcade positioned over the PIL, as well as remote brightenings at the footpoints of the fan surface linked magnetically to the breakout sheet, whereas the resistive-kink jet exhibits heating and/or nonthermal energy deposition in locations linked magnetically to the precessing reconnection site. In both scenarios, however, the bulk of the energy release is associated with the rapid reconnection between the twisted closed flux and the ambient open field, not with prior reconnection or with ideal processes. In the following Section, we use the breakout scenario to interpret the jet observations described in \S2.  

\section{RESULTS}\label{results}
\subsection{Magnetic field topology and evolution}
The breakout jet model provides a compelling framework for interpreting the evolving features of our observed jet. Figure \ref{hmi2}(a) displays a 3D view of the potential magnetic field extrapolated from an HMI magnetogram before the eruption ($\sim$16:40 UT). The mini-filament segment that erupted originally resided beneath the loops on the right side. We see many field lines connecting the central positive polarities to the surrounding negative-polarity regions, but only a single null point in the corona, as in our embedded-bipole jet model (\S \ref{model}). The open field lines of the coronal hole were rooted in the surrounding negative polarity regions (e.g., yellow lines in Figure \ref{hmi2}(a)). The initial plasma configuration (Fig. \ref{aia211}(b)) clearly traces this fan-and-spine topology: the fan surface was located outside the bright loops emanating from the positive polarity patch and terminating in the surrounding negative polarity region, while the outer spine was the spire dimly visible for hours before the jet. 

The existence of a mini-filament indicates that magnetic shear is concentrated at the PIL inside the fan, similar to large-scale filament channels. 
Figure \ref{hmi2}(b) displays a 3D top-down view of selected field lines in the vicinity of the eruption site, extrapolated from an HMI magnetogram 2 min before jet onset ($\sim$17:15 UT). The observed evolution of the magnetic field before and during the event, displayed in Figure \ref{hmi1} and the accompanying movie, reveals that little or no flux cancellation or emergence took place during the 16 hours before event onset. Therefore another mechanism must be invoked to explain the free energy buildup at the PIL that drove the eruption, as we discuss in \S \ref{conclusions}.

\subsection{Pre-jet activity: Slow reconnection}
To visualize the temporal evolution of the event, we chose a slice S1 through the axis of the rising circular feature (shown in Figure \ref{blobs}), and created time-distance (TD) intensity plots along S1 using AIA 304, 171, and 131 \AA~ images (Fig. \ref{st}(a-c)). Figure \ref{st}(a) shows the activation onset and slow rise of the CF and enclosed filament, starting at $\sim$16:45 UT (marked by the first vertical dotted line). The leading edge (LE) of the circular feature rose slowly until 17:11 UT, with a speed of $\sim$15 \kms (Fig. \ref{st}(a)). Multiple brightenings below the filament began at $\sim$16:53 UT and continued until 17:06 UT; each rose a short distance along S1 to roughly the same height. We interpret the brightenings that accompanied the activation and slow rise of the filament as signatures of magnetic reconnection beneath the filament, and the bright CF surrounding the filament as a flux rope formed by this reconnection.

TD plots along additional slices S2-S4, shown in Figure \ref{dim}, allowed us to detect and measure other dynamic features not immediately apparent in the full images. Figure \ref{st}(d)) reveals quasi-periodic moving features along S2, originating near the base of the outer spine, during the slow and fast rise phases (16:58 to 17:10 UT). These narrow features move along the same path at roughly the same constant speed, and contain very small, bright blobs (see movie accompanying Figure \ref{def}). The speed of one selected bright feature was $\sim$180$\pm$10 \kms, and the average period between features was $\sim$100 s. According to the breakout model, while the flux rope rises and expands slowly, the overlying closed flux gradually reconnects through the breakout sheet with the closed field on the other side of the null and with the external field bordering the dome. This slow breakout reconnection removes the restraining force holding down the flux rope and changes the connectivity of the overlying magnetic field, but very little free energy is lost in this phase because the twisted field in the flux rope is not yet involved. We interpret the inhomogeneous, narrow, quasi-periodic features as weak mass flows along the spine resulting from slow, bursty breakout reconnection. The intermittent, compact bright blobs are likely to be plasmoids, as seen in the breakout current sheet in our high-resolution simulations of large-scale CMEs \citep{karpen2012} and small-scale jets \citep{wyper2016b}. Because this reconnection is not intense, we do not detect remote brightenings at the footpoints of field lines that were processed through the breakout sheet. 

\subsection{Dimmings}
Figure \ref{st}(e) reveals the formation of two dimming regions, D1 and D2, during the activation and slow rise of the flux rope, starting at $\sim$16:48 UT for D1 and 16:53 UT for D2. In addition, as revealed by the TD plot along S4 (Fig. \ref{st}(f)), a strong elongated dimming region (D3) became visible at $\sim$16:50 UT and persisted throughout the observing period. The formation and growth of this dimming region, also visible in Figure \ref{dim} and the accompanying movie, indicates that the spine began to move when slow breakout reconnection commenced, and became more displaced from its initial position as the event progressed. The different onset times for D1 and D2 might be explained by a strong asymmetry between the two legs of the flux rope, leading to different expansion rates, but we cannot verify this with the available 2D images. 

\subsection{Plasmoid formation in flare current-sheet}
From $\sim$17:07 UT onward, especially during 17:12-17:14 UT, we detected multiple blobs in the bright, inverted-V shaped structure below the flux rope, along with the fast rise of the filament (Fig. \ref{blobs}). In Figure \ref{st}(b), boxes U and D encompass the upward and downward moving blobs, whose projected speeds are $\sim$135 and 55 \kms, respectively. Some blobs also appear to coalesce during their propagation. We attribute the growing linear features beneath the rising flux rope to plasma emission associated with a current sheet, analogous to the flare current sheet in CME/eruptive flare models \citep[e.g.,][]{karpen2012}. In this case, the multiple bright blobs are plasmoids formed by bursty reconnection in this current sheet, another phenomenon commonly found in high-Lundquist number reconnection simulations \citep[e.g.,][]{daughton2006,daughton2014,drake2006,fermo2010,uzdensky2010,huang2012,karpen2012,mei2012,cassak2013,guo2013,wyper2014b,wyper2014a,lynch2016,guidoni2016}. Multiple plasmoids moving bidirectionally were previously detected below flux ropes in active-region eruptive flares \citep{takasao2012,kumar2013,kumar2015}. If we assume a minimum base field strength of 50 G and an Alfv\'en speed of $\sim$135 \kms for an upward-moving plasmoid, we obtain an estimated minimum density of $4.5 \times 10^{10}$ cm$^{-3}$ for the flare current sheet.  The curious appearance of the bright, inverted-V shaped structure diverging beneath the flux rope (see Figure \ref{blobs} red and white arrows, Column 3, and the accompanying movie) underscores the 3D geometry of the flare current sheet.  Here the righthand bright line (marked by white arrows) appeared first ($\sim$17:05 UT), followed by the left one (marked by green arrow) at $\sim$17:13 UT.  The righthand line disappeared by $\sim$17:15 UT, while the left faded gradually through the rest of the observing period. A large downward-moving blob is visible during $\sim$17:17-17:18 UT. Because current sheets are very thin, they become visible only if the line of sight passes through multiple folds or through regions of enhanced density. We speculate that the appearance of two plasmoid-generating regions could be a sign of patchy reconnection in a rippled current sheet, with reconnection sites appearing at different locations along the sheet. 

\subsection{Formation of multiple brightenings and mini-flare arcade}
The abrupt change in dynamics starting around 17:12 UT coincided with the arrival of the flux-rope leading edge at the breakout current sheet. The filament, marked by F in Figure \ref{st}(a-b), accelerated rapidly to $\sim$126 \kms. At the same time, localized remote brightenings appeared; the brightest are labeled B1-B3 in Figure \ref{bright}(b). To establish the locations of these brightenings relative to the underlying magnetic structure, we performed a potential field extrapolation using the HMI magnetogram at 17:15 UT; the field of view is the same as in Figure \ref{hmi1}(d).  As is evident in Figure \ref{hmi2}(b), the two bright arcs labeled B1 appeared at the base of the flare current sheet, on either side of the PIL. Figure \ref{def} and the accompanying movie show that the longer righthand arc moved away from the left arc at a measured rate of $\sim$30 \kms for $\sim$3 min, after which the B1 arcs fade gradually until the end of the observation. Based on their locations on either side of the PIL and the progressive displacement of the right arc away from the PIL, we interpret B1 as ``mini-flare ribbons" whose apparent separation was limited spatially (particularly for the left ribbon) by the compact geometry within the dome. The movie accompanying Figure \ref{aia1} reveals that the post-flare arcade associated with the eruption was unimpressive: a feature resembling a bright loop connected the ribbons (Fig. \ref{jet}), but no progression or expansion along the PIL is visible in the hot channels. 

The B2 brightenings appeared near the footpoints of the bright, low-lying loops connecting the positive-polarity patch to the surrounding negative-polarity region (Fig. \ref{hmi2}(b)).  Therefore, we infer that the long, narrow, bright arc of B2 was located at the base of one side of the separatrix dome, at the footpoints of field lines passing through the breakout current sheet, with perhaps a fainter continuation on the other side (see rightmost green arrow in Fig. \ref{bright}(b)).  Similar to flare ribbons, such footpoint brightenings might be produced by accelerated electrons precipitating from the breakout reconnection region. Observations of analogous remote brightenings in association with breakout CMEs were reported by, e.g., \citet{sterling2001} and \citet{kumar2016,kumar2017}. Because projection effects make it difficult to discern whether the source was near the surface or higher in the corona, however, some of the B2 emission also could have come from hot, dense plasma in the breakout sheet or compressed portions of the nearby dome. In particular, the fainter B2 arc on the right side of the spine appears to be coincident with the outer edge of the erupting flux rope, as we discuss below. 

The bright feature B3 rose with the flux rope, and possibly results from compression and/or reconnection-associated heating.  The AIA 304 \AA~ mean counts within a box covering the source region (shown in Fig. \ref{bright}(a)) are overplotted as the red curve in Figure \ref{st}(c). The intensity at B3 peaked around 17:16 UT, when the top surface of the flux rope encounters the dome near the spine, where we expect the apex of the breakout sheet to reside.  The GOES soft X-ray flux profile in this interval was contaminated by simultaneous flare activity in a southern-hemisphere active region, and cannot be used here to infer the timing of the mini-flare emission. Therefore we use the light curve of the AIA 94 \AA~ mean counts (red curve in Figure \ref{st}(f)), extracted from same box shown in Fig. \ref{bright}(a), as a proxy for the mini-flare intensity.  The AIA 304 and 94 \AA~ mean count profiles suggest that fast flare reconnection beneath the flux rope coincided with fast breakout reconnection between the flux rope and the adjacent open field. The spatial distribution of distinct bands of increased emissions (B1-B3) signifies that both flare and breakout reconnection produced bulk heating and/or particle acceleration at this stage.  During the fast rise of the flux rope ($\sim$17:14-17:16 UT), the bright spine was deflected leftward and the leading edge of the flux rope pushed against the dome near the outer spine, causing a distinct bulge (Fig. \ref{def}). The simultaneous observation of strong deflection/displacement and an increase in overall high-temperature emissions (red curve in Figure \ref{st}(f)) suggest a feedback loop between the fast breakout reconnection and flare reconnection, accompanied by particle acceleration in or near both current sheets.  

\subsection{Untwisting jet}
\begin{figure*}
\centering{
\includegraphics[width=8.7cm]{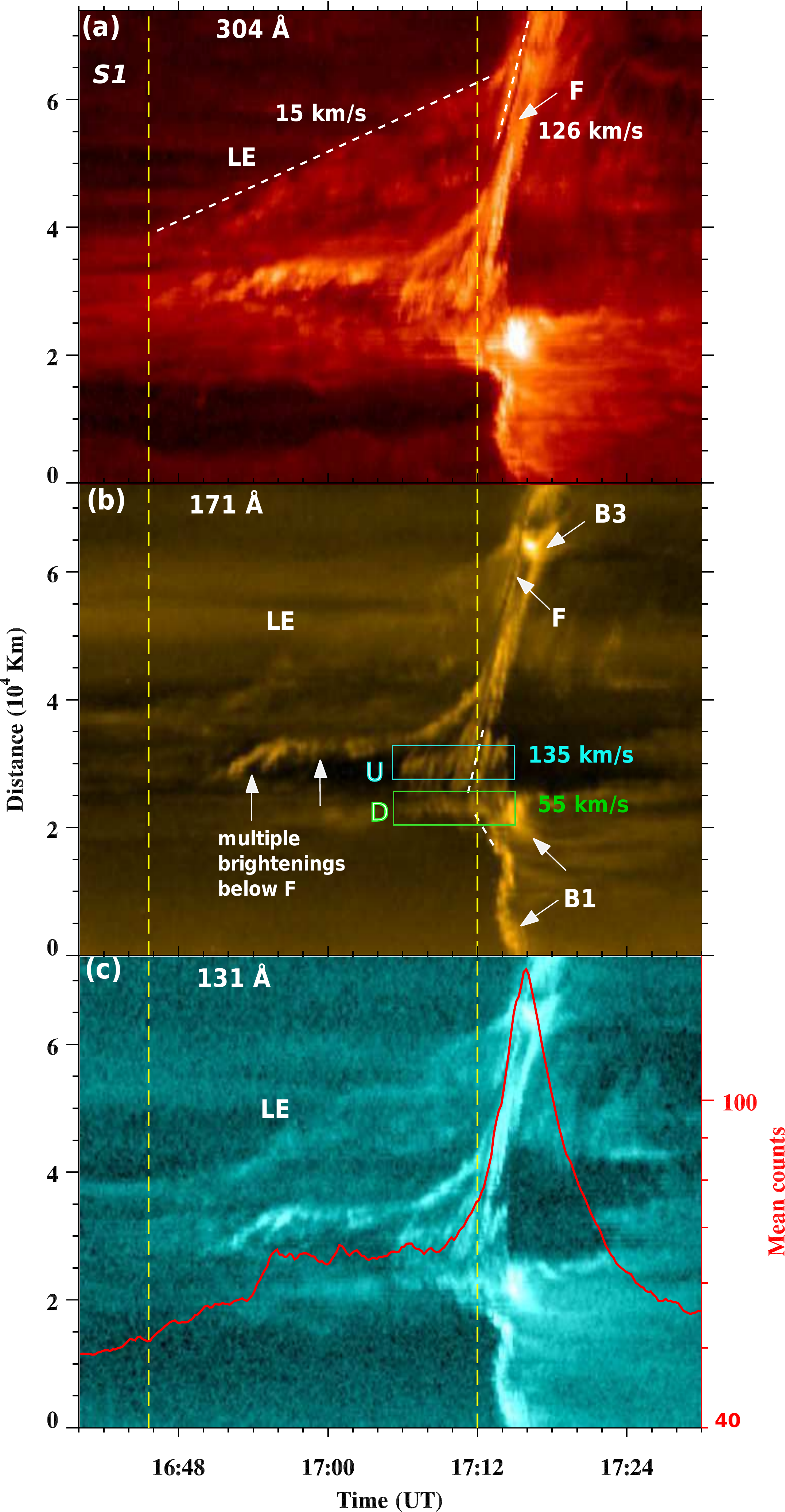}
\includegraphics[width=8.3cm]{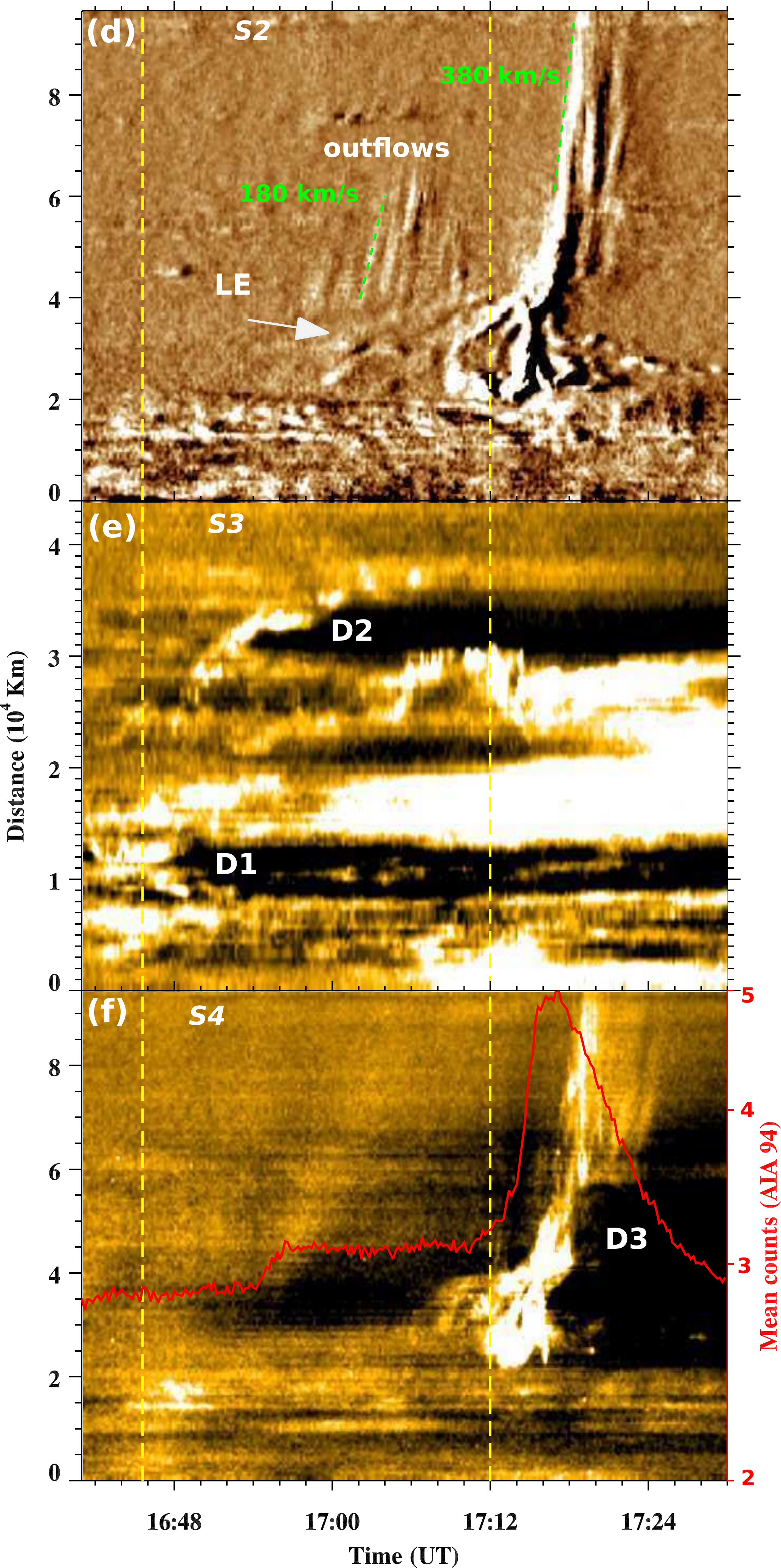}
}
\caption{(a-c) AIA 304, 171, and 131 \AA~ time-distance intensity plots along slice S1 (see Figure \ref{blobs}). LE is the leading edge of the rising flux rope, F is the dark filament, and boxes U and D outline upward and downward moving plasma blobs. The red curve in (c) is the mean counts extracted from AIA 304 \AA~ intensity images of the jet source region. The two vertical dashed lines (yellow) indicate the times of filament activation (16:48 UT) and onset of fast breakout reconnection (17:12 UT). (d-f) Time-distance intensity plots along slices S2, S3, and S4 (defined in Figure \ref{dim}) from (d) AIA 193 \AA~ (running difference) and (e,f) 171 \AA~ (base difference) images. 
} 
\label{st}
\end{figure*}

The most dynamic symptom of explosive reconnection was the expulsion of the untwisting jet (Fig. \ref{jet}). At 17:17 UT, the speed of the jet measured from Figure \ref{st}(a) was $\sim$380$\pm$20 \kms. The AIA movies accompanying Figures \ref{aia1} and \ref{blobs} clearly show the flux rope being destroyed by breakout reconnection, releasing a curtain of filamentary, multithermal plasma onto the external open field of the coronal hole. In fact, the portion of the flux rope between the footpoint rooted at D2 and the breakout site is quite visible around (175$\arcsec$,430-480$\arcsec$) in all 3 channels toward the end of the movie accompanying Fig. \ref{blobs}. This offers a striking example of interchange reconnection, in which the flux rope effectively exchanges footpoints with open flux rooted on the other side of the dome. The newly closed flux then must be rooted at D1 at one end and in negative polarity inside the separatrix at the other end; the most likely visible manifestation is a compact set of bright loops around coordinate (120$\arcsec$,400$\arcsec$) in the base-difference images of the movie accompanying Figure \ref{dim}. The coronal-temperature images show dark threads of absorbing material being flattened against the breakout sheet and subsequently propagating outward next to hot plasma, demonstrating that parts of the mini-filament remained cool even through reconnection.  After jet onset, Figure \ref{st}(f) shows that dimming region D3 expanded significantly, although it was partially obscured by the passage of the hot jet material. This expansion probably was caused by the deflection of nearby open field away from the spine, and by the evacuation of the preexisting mass in the Alfv\'en wave and mass flow of the jet. 

Due to projection effects and interference from other bright features along the line of sight, it was challenging to determine the direction in which the jet outflow rotated as it traveled outward (see movie accompanying Figure \ref{aia1}). After fast reconnection between the flux rope and the external field began, the flux rope appears to be rotating clockwise. As the flux rope opened up, the bright left edge transitioned from an arc to a linear feature aligned with the spine, while the contents appeared as a broad collection of threads parallel to the spine.  Thereafter the jet motion appears to be counterclockwise as viewed from above, with the threads apparently moving left to right as they rose. 

After the jet traveled out of the AIA field of view, the system started to relax back toward a lower energy state. The AIA 193 \AA~ base-difference image at 17:28 UT (Fig. \ref{dim}(a)) shows a cusp-shaped structure at the base of the jet, i.e., the apex of the separatrix dome. The southern portion of the mini-filament remained dynamic but did not erupt until the next day. \\   

The time sequence of the activity is summarized below:\\
{\small
\begin{tabularx}{\linewidth}{lX}
\hline \hline \\
16:45  & Mini-filament activation begins.\\
16:48 & Dimming region D1 appears at the southern end of the sigmoid.\\
16:53  & Brightenings below the filament start, signalling the initial formation of the flux rope, and dimming region D2 appears at the northern end of the sigmoid.\\
16:53--17:07  & Slow rise ($\sim$15 \kms) of the flux rope and filament.\\
16:58--17:10  & Quasi-periodic narrow outflows ($\sim$180 \kms) are expelled from the breakout current sheet while the flux rope rises. \\ 
17:07--17:12   & Fast rise ($\sim$126 \kms) of the flux rope and filament.\\
17:07-- 17:16 & Multiple bright plasmoids propagate up and down the flare current sheet below the flux rope.\\
17:12--17:16  & The flux rope encounters the bright dome at the breakout sheet, and intense surface and coronal brightenings (B1, B2, B3) appear.\\
17:12  onward & A strong elongated dimming region (D3) appears and expands to the left of the jet axis.\\
17:17--17:23  & Fast coronal jet propagates outward ($\sim$380 \kms).\\
\hline 
\end{tabularx}
}

\section{CONCLUSIONS}\label{conclusions}
We studied an on-disk coronal-hole jet associated with a mini-filament eruption on 9 January 2014. The HMI magnetograms do not exhibit measurable flux emergence or cancellation below the mini-filament channel for at least 16 hours before the jet, so we conclude that this mini-filament eruption was not directly powered or triggered by either mechanism. Both a potential-field extrapolation and the EUV images suggest that the pre-eruption magnetic configuration was an embedded bipole, consistent with our model for reconnection-driven coronal jets summarized in \S \ref{model} \citep{antiochos1990,antiochos1996,pariat2009,pariat2010,pariat2015,pariat2016,wyper2016a,wyper2016b,karpen2017}. In particular, the presence of the mini-filament many hours before jet onset confirms that the source region contained a highly sheared PIL, which is a requirement for our breakout jet model \citep{wyper2017,wyper2018}. The observations alone do not reveal how free energy built up at the PIL, but the absence of evidence for flux cancellation or emergence associated with the observed event points to coherent footpoint motions (as suggested by the HMI magnetograms) or helicity condensation \citep{antiochos2013} as possible candidates. Preliminary results from simulations of helicity condensation within an embedded bipole demonstrate that this mechanism can create a filament channel that subsequently erupts as a jet (C. DeVore, in preparation). 

In the idealized case of a symmetric magnetic-field distribution around the central minority polarity and uniform footpoint displacements, the filament channel forms more or less uniformly along the circular PIL (e.g., through helicity condensation, \citealt{knizhnik2015} ).  Under more realistic circumstances, however, certain locations are most likely to accumulate sufficient stress to drive an eruption. For example, the embedded bipole should have a concentration of majority polarity near the minority polarity peak, as in the \citet{wyper2017,wyper2018} numerical simulation, and the footpoint motions are probably nonuniform. In that case, the same photospheric motions would produce stronger magnetic shears at the PIL between the two concentrations than elsewhere, predisposing this portion of the channel to rise, form a flux rope, and erupt. We also speculate that the observed reconfiguration/rotation of the central minority polarity (positive) assisted in the formation of the sheared arcade hosting the filament and the subsequent eruption. The rotation was roughly clockwise as seen from above, which appears consistent with the filament orientation, but it is unclear why only the northern portion of the filament (which is not aligned with the positive patch) erupted.

The observed and modeled flux rope underwent three phases of evolution: slow rise, fast rise, and explosive eruption. The filament activation and slow rise were accompanied by small brightenings, likely due to slow reconnection below that formed a flux rope around the filament. The bright outline of the flux rope in the EUV images indicates that the plasma on the newly reconnected field lines was heated or compressed (or both) as the flux rope grew. Beneath the rising flux rope we observed thin, bright, linear features interpreted as different views of a 3D current sheet, which grew in length and produced bright dynamic blobs. To our knowledge, this is the first reported observation of multiple plasmoids moving bidirectionally during the build-up to a CH jet.  The upward moving plasmoids expanded and merged with the rising flux rope, while the downward-directed plasmoids created a modest mini-flare arcade that did not expand significantly. Similar bidirectional streaming of tearing-generated plasmoids and small flux ropes away from the separatrix surface was detected in the high-resolution simulations of resistive-kink jets \citep{wyper2016b}. In our observed jet, the plasmoid speeds were lower than those observed in larger-scale eruptions in active regions \citep{kumar2013}; this is not surprising, because the magnetic-field strength is significantly lower in these coronal-hole embedded bipoles. Our high-resolution numerical simulations of the breakout model for CMEs/solar eruptions \citep{karpen2012,guidoni2016} also manifest multiple plasmoids in the flare current sheet below the rising flux rope, and in the breakout current sheet above the flux rope.    

As in the breakout model for CMEs and eruptive flares \citep{antiochos1998,antiochos1999}, reconnection plays two roles in this CH jet: removal of the overlying restraining flux (breakout reconnection) and disconnection of the flux rope (flare reconnection). When the flux-rope leading edge arrived at the top of the separatrix between the closed and open flux systems, the outline of the flux rope and the adjacent separatrix surface brightened appreciably --- possibly a signature of heating by the onset of fast breakout reconnection. As seen in the simulations by \citet{wyper2017,wyper2018}, the flux rope was opened and destroyed by breakout reconnection, enabling both prominence and coronal plasma to escape as an untwisting jet.  Hot flux ropes (visible in AIA 131/94 \AA~ channels) have been observed previously during CMEs/eruptive flares that were consistent with the magnetic breakout model \citep{kumar2013fr,yurchyshyn2015}. Here, the erupting flux rope was observed in hot (131, 94 \AA) and cool (304, 171, 193 \AA) AIA channels. 

Three dimming regions formed during the observed event. A pair of EUV dimming regions was generated at the ends of the gradually rising flux rope as it drove slow breakout reconnection, and persisted during and after the explosive eruption. Similar dimmings associated with CMEs \citep[e.g., ][]{manoharan1996,sterling1997,thompson2000,miklenic2011,mason2014} have been interpreted as density depletions resulting from the opening of previously closed magnetic fields. A similar explanation applies equally well to this coronal breakout jet: the depletions began as the flux rope rose and its length increased, and continued as the plasma in the reconnecting flux rope was ejected onto open CH flux. The third dimming region originated next to the initial spine well before eruption and grew rapidly in angular extent at the start of fast breakout reconnection. This narrow, elongated dimming is a signature of the displacement of the dense spine through slow and fast breakout reconnection, the evacuation of the narrow jet channel by the ejecta and wave, and the deflection of nearby open field when the jet was triggered.  

Both qualitative and quantitative comparisons between the observed event and the simulated breakout jet demonstrate remarkable agreement.  The initial configuration inferred from the observations closely matches the embedded-bipole magnetic topology, with its fan, spine, and null point.  The origin of the highly sheared PIL in both cases cannot be attributed to flux cancellation or emergence, but rotational motions may play a role. Only half of the mini-filament erupted in both the observed and simulated events, leaving the other half to erupt later. During the flux rope's slow rise, both the observations and the simulations exhibit weak, repetitive outflows well above the flux rope, as well as progressive displacement of the spine, attributable to slow breakout reconnection. Multiple plasmoids move up and down in the lengthening current sheet below the observed and simulated flux ropes.  In both cases, explosive activity occurs only when the flux rope reaches the breakout sheet and triggers fast reconnection there.  The associated formation of mini-flare ribbons, remote ribbons connected to the breakout sheet, and the mini-flare arcade occur in the same sequence and in analogous locations in the observed and modeled events. The expulsion of a helical, trans-Alfv\'enic jet containing both ambient coronal and filament-channel plasma characterizes both the observed and simulated events. 

In the breakout jet simulation, the activity (slow rise, fast rise of flux rope, onset of fast breakout, and jet phase) persists for $\sim$40 min, which is comparable to the duration of our observed event. In addition, the flux rope slow- and fast-rise profiles are consistent with the flux rope LE kinematics in our observation. The measured jet outflow speed ($\sim$380 \kms) is also comparable with the Alfv\'enic outflow speed ($\sim$300 \kms) in the simulation, and consistent with the average speed of X-ray jets observed by Hinode XRT \citep{cirtain2007}. The quantitative results of the simulations scale with the assumed physical conditions, however, so these points of agreement should be viewed as encouraging rather than definitive.
 
In addition, we also remark that the breakout model can be applied to homologous jets. First, we note that the sheared system never loses all of its shear. In fact, the observed and simulated jets both involved a partial filament eruption; for the observed jet, the rest of the filament erupted a day later. This is common for CMEs as well. So one way to obtain recurrent jets is to invoke sequential partial eruptions. Second, as we found for the resistive-kink jets, continued driving reforms the filament channel after the first eruption. The timescale is uncertain, though, because the driving in these simulations is much faster than observed speeds. With some simple assumptions, we estimate the reformation time as follows. The filament channel formed approximately 18 min into the simulation with a driving speed peaking around 30 \kms. If we assume the channel would form more slowly under realistic solar conditions, by an amount equal to the ratio of the actual to the simulated driving speeds, then a typical observed photospheric flow speed of 1.5 \kms would produce a new filament channel in around 360 min (6 hours). For homologous eruptions this time gap is an upper bound, because the shear left in the filament channel after the first eruption should enable the channel to reform more quickly.

In conclusion, we report an outstanding example of a CH jet that was triggered by fast breakout reconnection above and flare reconnection below a filament-containing flux rope. The observation supports our breakout model for CH jets, and demonstrates that neither flux emergence nor cancellation is required to power or trigger these events. Ongoing analysis of a larger sample of observed CH jets in equatorial coronal holes is expected to shed more light on the underlying physical mechanisms. We also look forward to learning more about CH jet properties in the outer corona and beyond from the upcoming Solar Orbiter and Parker Solar Probe missions. 

\acknowledgments
We thank N. Raouafi and V. Uritsky for stimulating discussions. SDO is a mission for NASA's Living With a Star (LWS) program. This research was supported by PK's appointment to the NASA Postdoctoral Program at the Goddard Space Flight Center, administered by the Universities Space Research Association through a contract with NASA, and by a grant from the NASA Heliophysics Supporting Research program. PFW was supported through an award of a Royal Astronomical Society Fellowship. Figure 11 was produced by VAPOR (www.vapor.ucar.edu), a product of the Computational Information Systems Laboratory at the National Center for Atmospheric Research.


\bibliographystyle{apj}
\bibliography{reference.bib}

\end{document}